\documentclass[journal=jacsat,manuscript=article]{achemso}
\setkeys{acs}{keywords = true}
\setkeys{acs}{abbreviations= true}
\usepackage{chemformula} 
\usepackage[T1]{fontenc} 
\usepackage{xcolor}
\usepackage{graphicx}
\usepackage{subcaption}
\usepackage{pgfplots}
\usepackage{hyperref}
\usepackage{amsmath}
\usepackage{csquotes}
\usepackage[inline]{enumitem}
\usepackage{lineno}
\usepackage{tikz}
\usetikzlibrary{arrows}
\usetikzlibrary{positioning}
\usepackage{mathrsfs}
\usepackage{algorithm}
\usepackage{algpseudocode}
\usepackage{fancyvrb}

\SectionNumbersOn

\hypersetup{
  colorlinks,
  citecolor=violet,
  linkcolor=red,
  urlcolor=blue}

\newcommand*{\ie}{i.e.}

\newcommand*{\eg}{e.g.}
\newcommand*{\fig }{Fig.}

\newcommand*{\hsd}{Hofmeister dataset}

\newcommand*{\hz}{Hz}
\newcommand*{\khz}{kHz}

\newcommand*{\s}{\xi}
\newcommand*{\ft}{\mathscr{F}}
\newcommand*{\fm}{f_\mathrm{mean}}

\DeclareMathOperator{\sinc}{sinc}

\author{Jeyashree Krishnan}
\affiliation{Joint Research Center for Computational Biomedicine, RWTH Aachen University, Germany}
\email{krishnan@aices.rwth-aachen.de}
\author{Zeyu Lian}
\affiliation{Joint Research Center for Computational Biomedicine, RWTH Aachen University, Germany}
\email{zeyu.lian@rwth-aachen.de}
\author{Pieter E. Oomen}
\affiliation{Department of Chemistry and Molecular Biology, University of Gothenburg, Sweden
}
\email{pieteroomen@gmail.com}
\author{Xiulan He}
\affiliation{Department of Chemistry and Molecular Biology, University of Gothenburg, Sweden
}
\email{xiulan.he@gu.se}
\author{Soodabeh Majdi}
\affiliation{Department of Chemistry and Molecular Biology, University of Gothenburg, Sweden
}
\email{s_majdi59@yahoo.com}

\author{Andreas Schuppert}
\affiliation{Joint Research Center for Computational Biomedicine, RWTH Aachen University, Germany}
\email{schuppert@aices.rwth-aachen.de}

\author{Andrew Ewing}
\affiliation{Department of Chemistry and Molecular Biology, University of Gothenburg, Sweden
}
\email{andrew.ewing@gu.se}
\title[Spike-based Frequency Analysis]
  {Spike-by-Spike Frequency Analysis of Amperometry Traces Provides Statistical Validation of Observations in the Time Domain}

\abbreviations{FFT, IVIEC, SCA, PSF, DFT, EEG, MEG, EMG, fMCI, DTI, DMSO, GUI}
\begin{document}

\clearpage
\begin{abstract}
Amperometry is a commonly used electrochemical method for studying the process of exocytosis in real-time. Given the high precision of recording that amperometry procedures offer, the volume of data generated can span over several hundreds of megabytes to a few gigabytes and therefore necessitates systematic and reproducible methods for analysis. Though the spike characteristics of amperometry traces in the time domain hold information about the dynamics of exocytosis, these biochemical signals are, more often than not, characterized by time-varying signal properties. Such signals with time-variant properties may occur at different frequencies and therefore analyzing them in the frequency domain may provide statistical validation for observations already established in the time domain. This necessitates the use of time-variant, frequency-selective signal processing methods as well, which can adeptly quantify the dominant or mean frequencies in the signal. The Fast Fourier Transform (FFT) is a well-established computational tool that is commonly used to find the frequency components of a signal buried in noise. In this work, we outline a method for spike-based frequency analysis of amperometry traces using FFT that also provides statistical validation of observations on spike characteristics in the time domain. We demonstrate the method by utilizing simulated signals and by subsequently testing it on diverse amperometry datasets generated from different experiments with various chemical stimulations. To our knowledge, this is the first fully automated open-source tool available dedicated to the analysis of spikes extracted from amperometry signals in the frequency domain.

\end{abstract}

\clearpage
\label{sec:intro}

\begin{figure}[!htbp]
  \centering
  \includegraphics[scale=0.7]{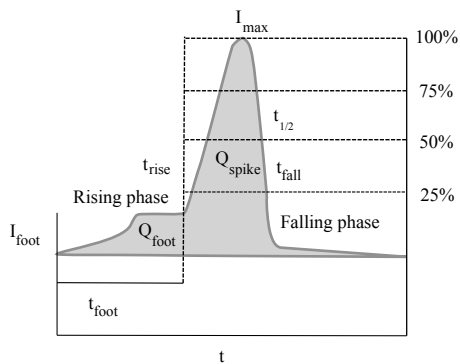}
  \caption{Spike parameters in the time domain: An amperometric spike is characterized by the Pre-Spike Foot (PSF), rising phase, spike and the falling phase. Here $I_\mathrm{max}$ is the peak current, $t_\mathrm{rise}$ is the rise time (from $25\%$ to $75\%$ of $I_\mathrm{max}$), $t_\frac{1}{2}$ is the half peak width, $I_\mathrm{foot}$ is the PSF current, $t_\mathrm{foot}$ is the PSF duration, $Q_\mathrm{spike}$ is the charge of the spike and $Q_\mathrm{foot}$ is the charge of the PSF.}
  \label{fig:spike}
\end{figure}

Amperometry is a commonly used method for studying the process of exocytosis in real time. It is useful in analyzing exocytosis because it offers high sensitivity, excellent temporal resolution, precise quantification of released neurotransmitters, and enables direct observation of the kinetics of secretory events as it happens \cite{Wrenn2003, Liu2019, Colliver2000}. Amperometric traces are generated by the oxidation of catecholamines released by a cell close to the microelectrode tip. The exocytosis process has been well-studied and typically progresses in the following molecular steps: \begin{enumerate*} \item Opening of the fusion pore resulting in the detection of the Pre-Spike Foot or PSF (indicated by the foot parameters $I_{\mathrm{foot}}$, $Q_{\mathrm{foot}}$, $t_{\mathrm{foot}}$ in \fig\ \ref{fig:spike}) \item Expansion of the fusion pore resulting in  massive release of catecholamines shown as a spike in the amperometric recording (steep rising phase characterized by $t_{\mathrm{rise}}$ in \fig\ \ref{fig:spike}) \item A decay phase caused by the pore closure ((double) exponential falling phase characterized by $t_{\mathrm{fall}}$ in \fig\ \ref{fig:spike})\end{enumerate*}. The frequency and shape of amperometry spikes and their sequence contain information about the dynamics of the release process, while their areas correspond to the total charge or the number of molecules released \cite{Segura2000, mosharov2005analysis}. Statistical analysis of these spikes can offer insights into patterns and correlations across different samples, with high reliability and generalisability for relatively modest time and resource costs. 

The microelectrode material is chosen based on it allowing high-throughput fabrication, fine spatial resolutions and fast analysis rates for single-cell analysis - most often this is carbon \cite{Lemaitre2014}.  Spikes generated during the exocytosis process may contain information about cell types or activity states of the cell. These spikes exhibit large temporal variability, and also depend on the type of stimulation used to induce exocytosis by the cell. Usually, the falling phase of a spike can be fitted by a single or double exponential function due to the diffusion mechanism. Quantification of electroactive neurotransmitters (\eg\ catecholamines) can be accomplished using different techniques based on amperometry. Single cell amperometry (SCA) was first introduced by Wightman and colleagues in the 1990s \cite{Wightman1991}. It employs a carbon microelectrode that is placed on a single cell in buffer with the assistance of a microscope. A reference electrode is located nearby in the same solution. The cell is (chemically) stimulated, after which it releases neurotransmitters from vesicles docked at the cell membrane. Since a constant potential is applied, the neurotransmitters are oxidized at the working electrode, leading to current \enquote{spikes} when this data is plotted versus time. Integration of these signals gives the charge Q, which, using Faraday’s law, allows quantification of the number of molecules involved in the \enquote{spike} \cite{Phan2017}. 


Another amperometric method, which does not measure the number of released molecules, but uses the same amperometric principle to quantify the contents of the vesicles that store neurotransmitters is VIEC (Vesicle Impact Electrochemical Cytometry). VIEC allows isolated vesicles to adsorb to a microelectrode which then burst stochastically, after which the electroactive contents are oxidized on the electrode surface. Similar to SCA, the number of molecules can then be quantified by examining the resulting current spikes \cite{Li2016, Dunevall2015}. Intracellular VIEC (IVIEC), employs a nanotip electrode that is inserted into a single cell \cite{Li2015}. This allows quantification of vesicular content inside a cell. By combining with SCA and comparing the number of molecules, different exocytotic modes can be investigated (e.g., kiss and run, partial release, and full release). Indeed, it has been found that exocytosis is a highly modulated process, and that partial release of vesicular content is preferred over all-or-nothing exocytosis both in vitro and in vivo \cite{Phan2017, Li2016, Ren2016, Larsson2020, Wang2021}. Vesicle impact electrochemical cytometry offers a direct quantification of vesicular catecholamine storage in isolated vesicles. VIEC is similar to IVIEC, but instead of in situ quantification of vesicle content in a living cell, vesicles are isolated and collected as a suspension in an intracellular physiological buffer to perform electrochemical cytometry \cite{Steven2019}.

Not unlike modern electrophysiological experiments, amperometry experiments also involve the acquisition, display, and analysis of data \cite{Wacker2013}. Given the high precision of recording amperometry procedures offer, the volume of data generated can span over several hundreds of megabytes to a few gigabytes. Several well-structured programs that facilitate the analysis of such data exist in the electrochemistry community, the most commonly used being IgorPro QuantaAnalysis software \cite{mosharov2005analysis}. QuantaAnalysis is a mature software dedicated to the analysis of amperometry traces that allows digital filtering and analysis of the current noise, spike identification, calculation of over $20$ spike kinetic parameters, and visualization. The most commonly measured spike characteristics include the area under the spike (such as $Q_\mathrm{spike}$ or $Q_\mathrm{foot}$), spike maximal height ($I_\mathrm{max}$) and spike width ($t_\mathrm{\frac{1}{2}}$) at half its height (where $I= \frac{I_{\mathrm{max}}}{2}$) as illustrated in \fig\ \ref{fig:spike}. 

Along with this, the QuantaAnalysis software outputs several other spike parameters in a matrix form (similar to that shown in \fig\ \ref{fig:workflow}(B)) that enables spike-wise or average characterization of specific traces. However, the QuantaAnalysis software involves semi-manual intervention for identifying thresholds for spikes extraction, selecting baseline intervals, and data manipulations in Excel. Furthermore, each amperometric trace has to be handled separately in QuantaAnalysis, which significantly affects the efficiency of the process. In addition, applications of neurochemistry to amperometry and their applications in single-cell analysis, is still a maturing area and because of the complex nature of the data, methods to analyze traces may vary considerably from lab to lab, thereby making it difficult to compare observations from different sources \cite{Evanko2005}. 

Further, though one can argue that the spike characteristics of amperometry traces in the time domain hold information about the dynamics of exocytosis, these biochemical signals are more often than not characterized by time-varying signal properties (\ie\ from the statistical perspective, they are non-stationary). Such signals with time-variant properties may occur at different frequencies and it may be difficult to catch subtle differences in their patterns by only using time-domain observations.  This necessitates the use of time-variant, frequency-selective signal processing methods as well, which can ably quantify the dominant or mean frequencies in the signal. The Fast Fourier Transform (FFT) is a well-established computational tool that is commonly used to find the frequency components of a signal buried in noise \cite{cooleyandtuckey1965, bracewell1999, Brigham1967, cohen1989, cooley1967, Cooley1969}. It is based on the Fourier Analysis method which states that any periodic function can be represented as an infinite enumerable sum of trigonometric functions \cite{Letelier2000}. 

FFT is a method for efficiently computing the Discrete Fourier Transform (DFT) of time series and facilitates power spectrum analysis and filter simulation of signals. All these measures are time-variant. Frequency analysis has found itself a wide variety of applications including digital image processing reconstruction, numerical solution of differential equations, multiple time series analysis, and filtering among many others \cite{Stankovic1994, alsopandnawroozi1966, Heitler2007, Mikheev2015,papoulis1962}. In the biomedical signal processing community, FFT is widely used in Electroencephalogram (EEG), Magnetoencephalography (MEG), EMG (Electromyogram), functional Multineuron Calcium Imaging (fMCI), analysis of calcium fluorescence traces, and Diffusion Tensor Imaging (DTI) data \cite{Tibau2013, Ruffinatti2011}. 

\begin{figure}[!htbp]
\centering
\includegraphics[scale=0.5]{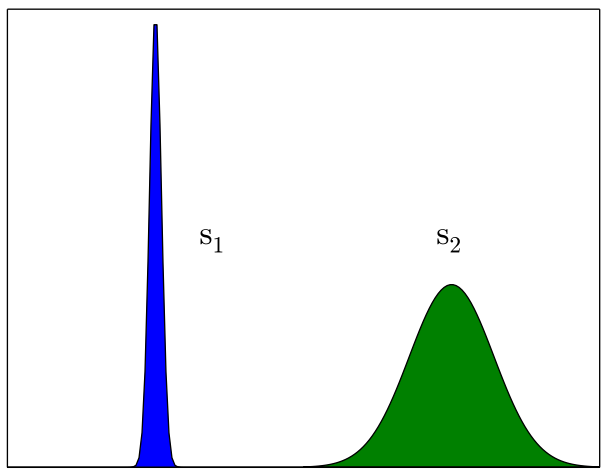}
\caption{Relationship between spike shape and mean frequency: Consider two spikes coming out of an amperometric trace, $s_1$representing a high frequency oscillation (hence a thin spike) and $s_2$ representing a low frequency oscillation (and hence a wide spike). We expect that the wider the curve, the lower is the mean frequency \ie\ $f_\mathrm{mean}(s_1) > f_\mathrm{mean}(s_2)$.}
\label{fig:hypothesis}
\end{figure}

In this work, we propose a method for spike-by-spike frequency analysis of amperometry traces and show that in addition to gaining information on the key frequencies, this method can provide statistical validation of observations made on traces in the time domain. We motivate this method through the relationship of spike width and mean frequency wherein thin spikes that arise out of high-frequency oscillations are expected to have higher mean frequency compared to their wider counterparts (such as that in the illustration \fig\ \ref{fig:hypothesis}). Amperometry traces show a variety of spike shapes intra-trace and inter-traces whose statistics usually remain consistent across a given stimulation and cell type. Hence it makes for a very interesting field of research for frequency analysis methods. In addition, the frequency analysis pipeline dedicated for amperometry traces that we implemented in Python is available open-source (refer to supplementary section). 

The paper is organized as follows: first, we motivate the idea of frequency analysis with simulated signals as a proof-of-concept, followed by outlining the key results for the aforementioned candidate datasets; then we summarize the work and give brief conclusions. Finally,  we go in-depth about the FFT and briefly describe the package implementation. Further details on the experimental methods for data generation, dataset attributes, and frequency analysis program may be found in the supplementary information section.
\section*{Results and discussion}
\label{sec:res}

\subsection*{Overview of the Datasets}
We demonstrate the method by utilizing simulated signals and subsequently test it on diverse amperometry datasets generated through different experiments under different stimulation conditions. Simulated signals or artificial spike trains were generated through spikes modeled with a linear rise and Gaussian decay. The three candidate datasets we chose to explore in the frequency domain are \begin{enumerate*} \item Hofmeister series dataset \cite{He}, \item Dimethyl Sulfoxide (DMSO) dataset \cite{Majdi2017}, and \item Electrodes dataset (first presented in this study) \end{enumerate*}. The Hofmeister series dataset makes an excellent candidate to demonstrate that the spike-by-spike frequency analysis method preserves time-domain spike characteristics. The investigation of the relationship between inorganic anions and exocytosis was carried out by He et. al. \cite{He} and it was shown that anions regulate pore geometry, opening duration, and pore closure in the exocytosis process. 

Specifically, when chromaffin cells were stimulated by couteranions of Hofmeister series ($\mathrm{Cl}^-$, $\mathrm{Br}^-$, $\mathrm{NO}_3^-$, $\mathrm{ClO}_4^-$, $\mathrm{SCN}^-$) in $\mathrm{K}^+$ solution, the spike width (including $t_\mathrm{rise}$, $t_\mathrm{\frac{1}{2}}$ and $t_\mathrm{fall}$) increases and the PSF parameters (including $N_\mathrm{molecules}$,  $\frac{N_\mathrm{foot}}{N_\mathrm{events}}$
and $I_\mathrm{foot}$ where $N$ is the number of molecules and $I$ is the current) decreases in the Hofmeister order while the number of spike events appeared to be similar across all stimulations. With the stimulation of chaotropic anions (such as $\mathrm{SCN}^-$), the expansion and closing time of the fusion pore is longer compared to that of kosmotropic ions (such as $\mathrm{Cl}^-$). The Hofmeister series dataset has therefore been well-studied in the time domain. 

Another compound, DMSO has also been shown to affect the fusion pore opening rate and increase neurotransmitter content while leaving vesicular contents unchanged. Unlike for example, the Hofmeister series, DMSO affects only the rising phase, $t_\mathrm{rise}$. The DMSO dataset was generated through IVIEC experiments conducted on chromaffin cells using a nanotip electrode. Analysis of control and $0.6\%$ DMSO datasets in the time domain has been established and hence makes an interesting dataset for frequency analysis. We also demonstrate frequency analysis methods on the electrode dataset which constitute VIEC experiments on chromaffin vesicles using three electrode materials: carbon, platinum, and gold. 

\subsection*{Hypothesis Verification using Simulated Signals }
To verify the hypothesis of the relationship between spike shape and mean frequency, we generated simulated signals that mimic the behavior of the Hofmeister ions in the time domain. Simulated to mimic the amperometry spikes, the artificial spikes consist of a linear rising segment and a Gaussian decay. Note that Gaussian decay models an exponentially decaying amperometry spike which simulates the signal decay more accurately compared to Dirac delta spikes. 

\begin{figure}[!htbp]
\centering
\textbf{(A)} \includegraphics[scale=0.4]{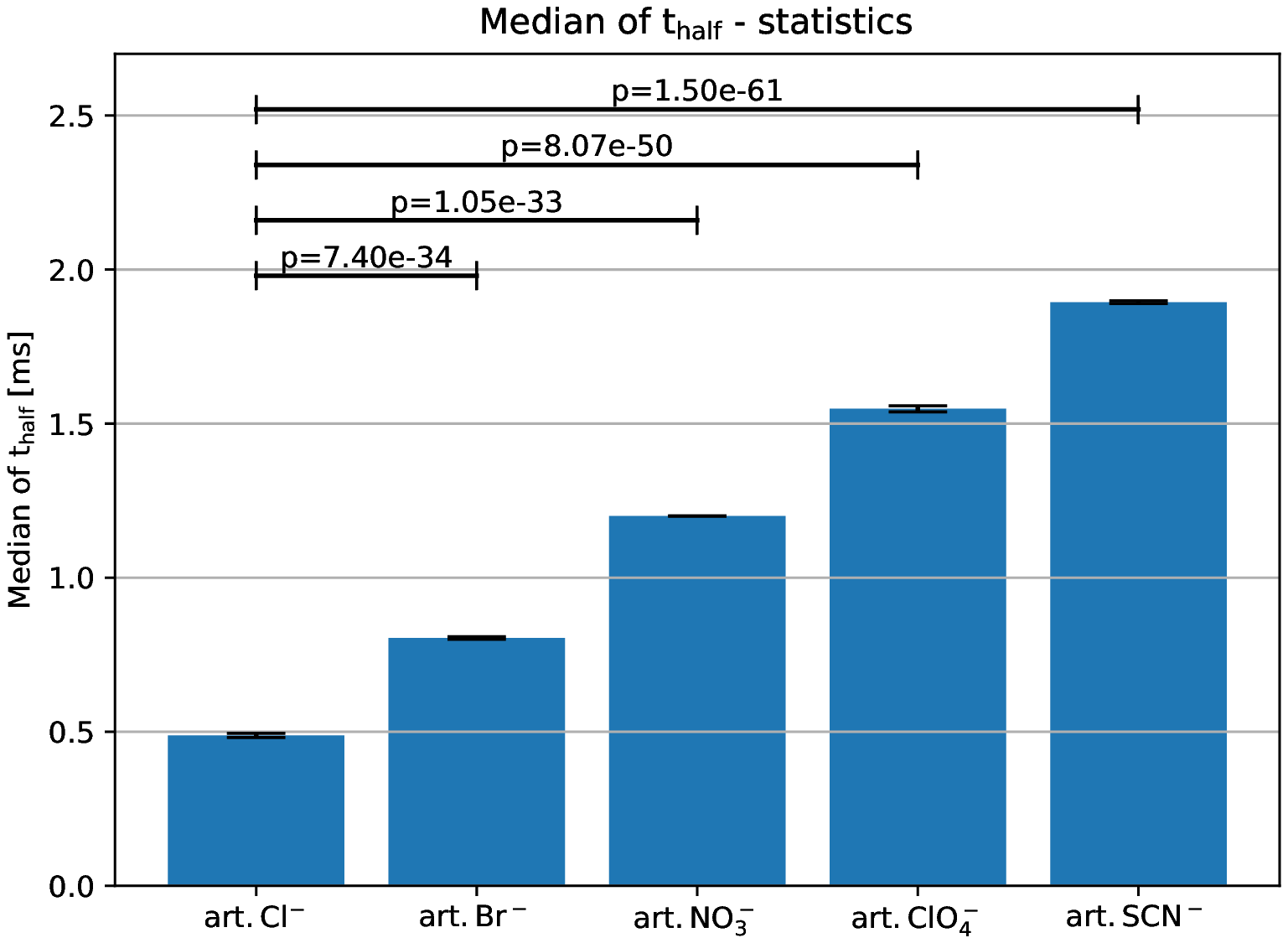}
\textbf{(B)} \includegraphics[scale=0.4]{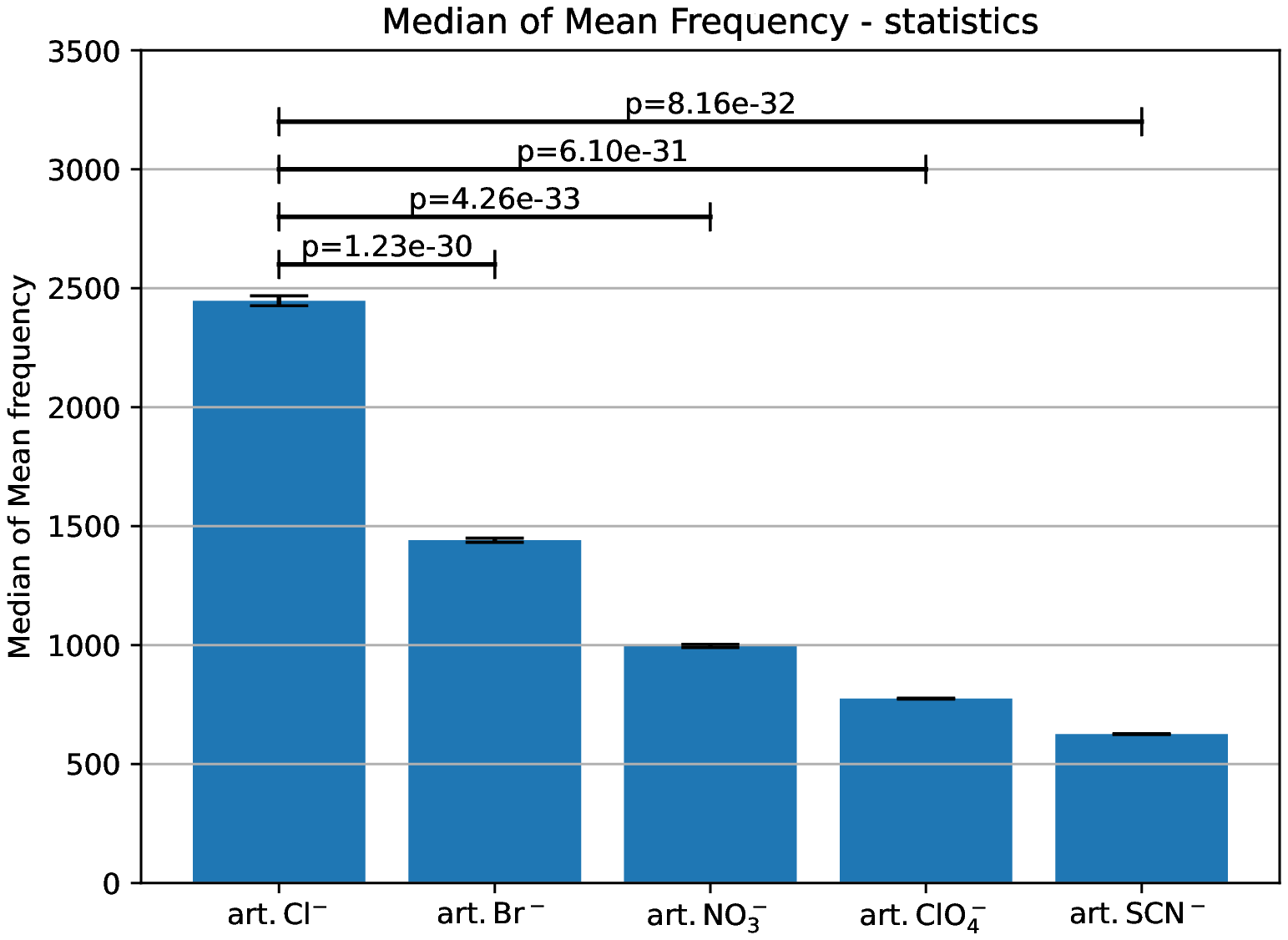}
\caption{FFT on simulated signals: (A) Spike characteristics in the time domain shown by median of $t_\mathrm{half}$. (B) Spike characteristics in the frequency domain shown by median of mean frequency, $f_\mathrm{mean}$. Hofmeister-like arrangement of mean frequency in the simulated spike trains (\enquote{art} stands for artificial). Standard error of mean of the median of $t_\mathrm{half}$ and the median of mean frequency are shown by the error bars.}
\label{fig:art_fmean}
\end{figure}

The artificially generated set of spike trains mimics the behaviors of the Hofmeister series dataset observed in the time domain, \eg\ kosmotropic anions in the Hofmeister series cause thin spikes (hence high-frequency oscillations) in comparison to the wide spikes (low-frequency oscillations) of the chaotropic ions. In other words, in the time domain, the median of the spike width, $t_\mathrm{\frac{1}{2}}$ of the artificial spikes increases in average along the \enquote{Hofmeister} order (see \fig\ \ref{fig:art_fmean}). 

Thus we artificially assigned each anion type in the artificial data with a range of width that does not overlap with the others, as can be seen in \fig\ \ref{fig:art_fmean}(A). Since the number of samples generated per artificial data category was sufficient and homogeneous across categories, we see that the width of standard error of mean bar is quite short and relatively consistent across all categories along with high confidence intervals. In addition, for a detailed description of artificial data generation procedure refer to the supplementary section. 

We found that the averaged mean frequency of the artificially generated spike trains decreases along the Hofmeister order with no exception, which is consistent with the observations on real data. Since a thinner sine function oscillates with a higher frequency, a kosmotropic anion like $\mathrm{Cl}^-$ will behave similarly, as can be seen from \fig\ \ref{fig:art_fmean}(B). 

\begin{figure}[!htbp]
\centering
\begin{subfigure}{.4\linewidth}
\includegraphics[scale=0.4]{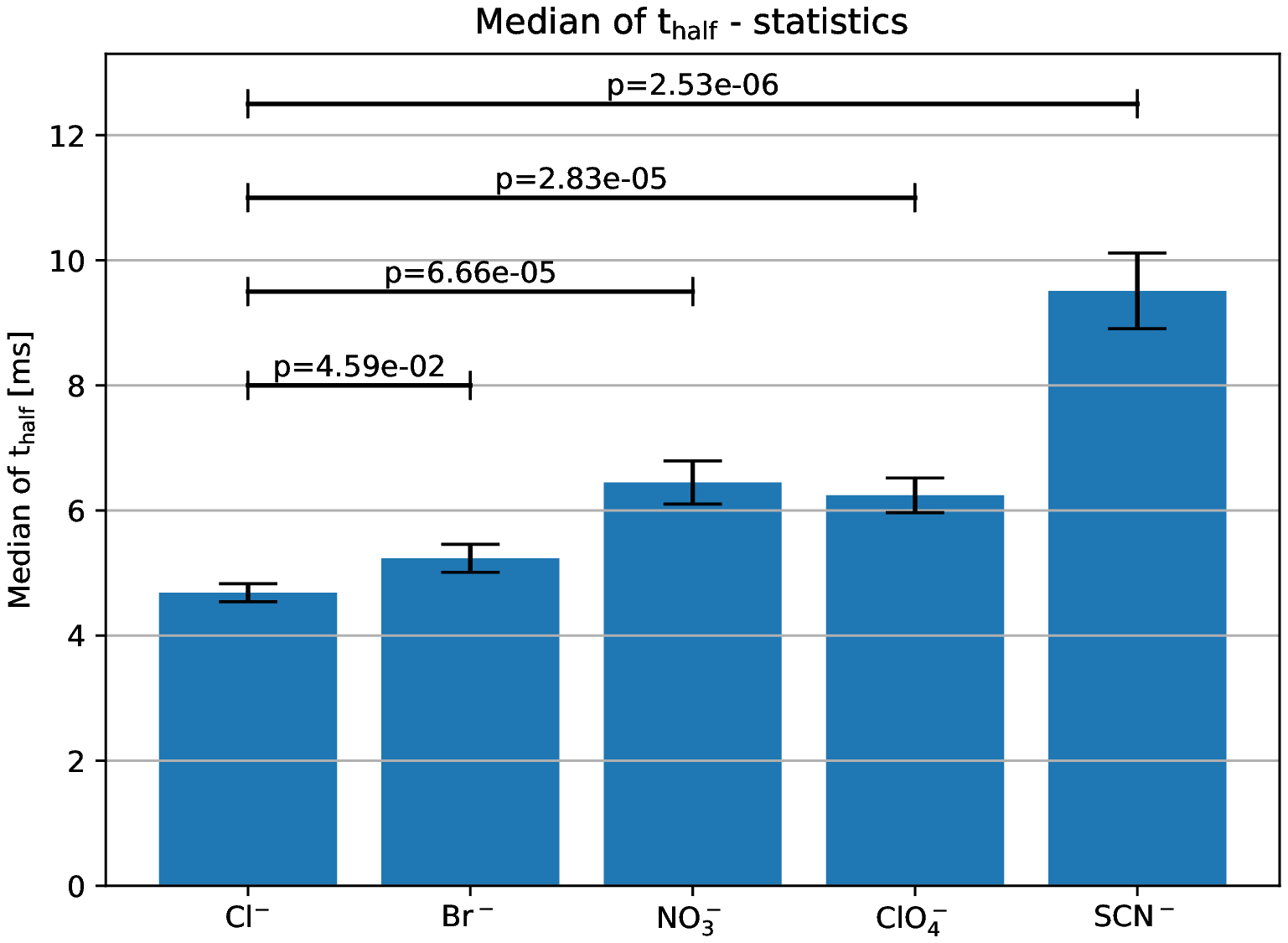} 
\end{subfigure}
\begin{subfigure}{.4\linewidth}
\includegraphics[scale=0.4]{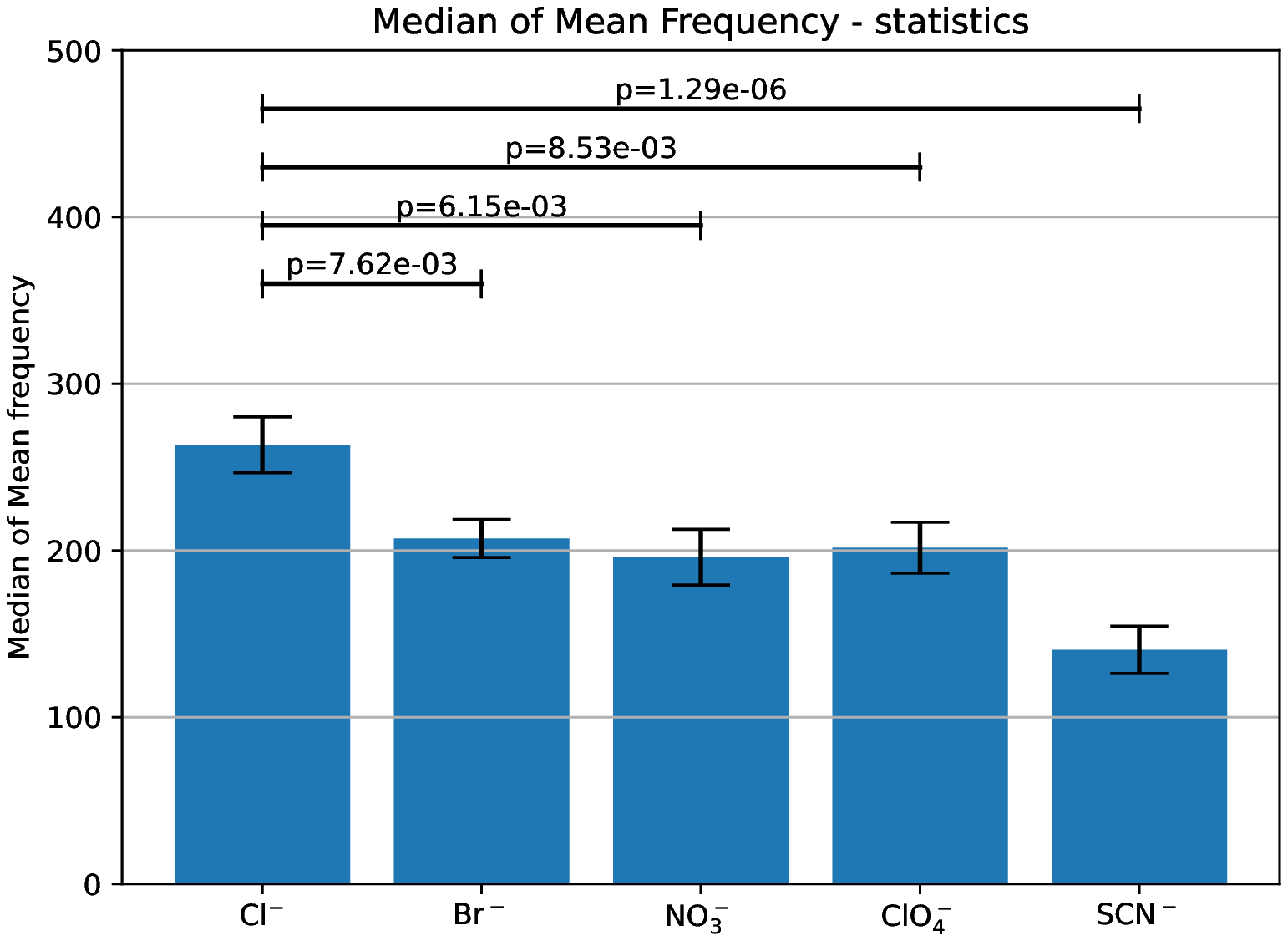}
\end{subfigure}
\caption{FFT on the Hofmeister dataset. Left: Spike characteristics due to different anion stimulation in the time domain shown by $t_\mathrm{half}$. Right: Spike characteristics in the frequency domain shown by median of mean frequency, $f_\mathrm{mean}$. The averaged median of mean frequency (shown with solid bar) shows a clear dependency on the stimulating anion, \ie\ it decreases along the Hofmeister-order (here from left to right) except for the anion $\mathrm{ClO_4^{-}}$, which also showed its abnormality during the time-domain analysis. Additionally, the cross-cell standard error of mean of the median of $t_\mathrm{half}$ and the median of mean frequency are denoted by the error bars.}
\label{fig:hof_fmean}

\begin{subfigure}{.4\linewidth}
\includegraphics[scale=0.4]{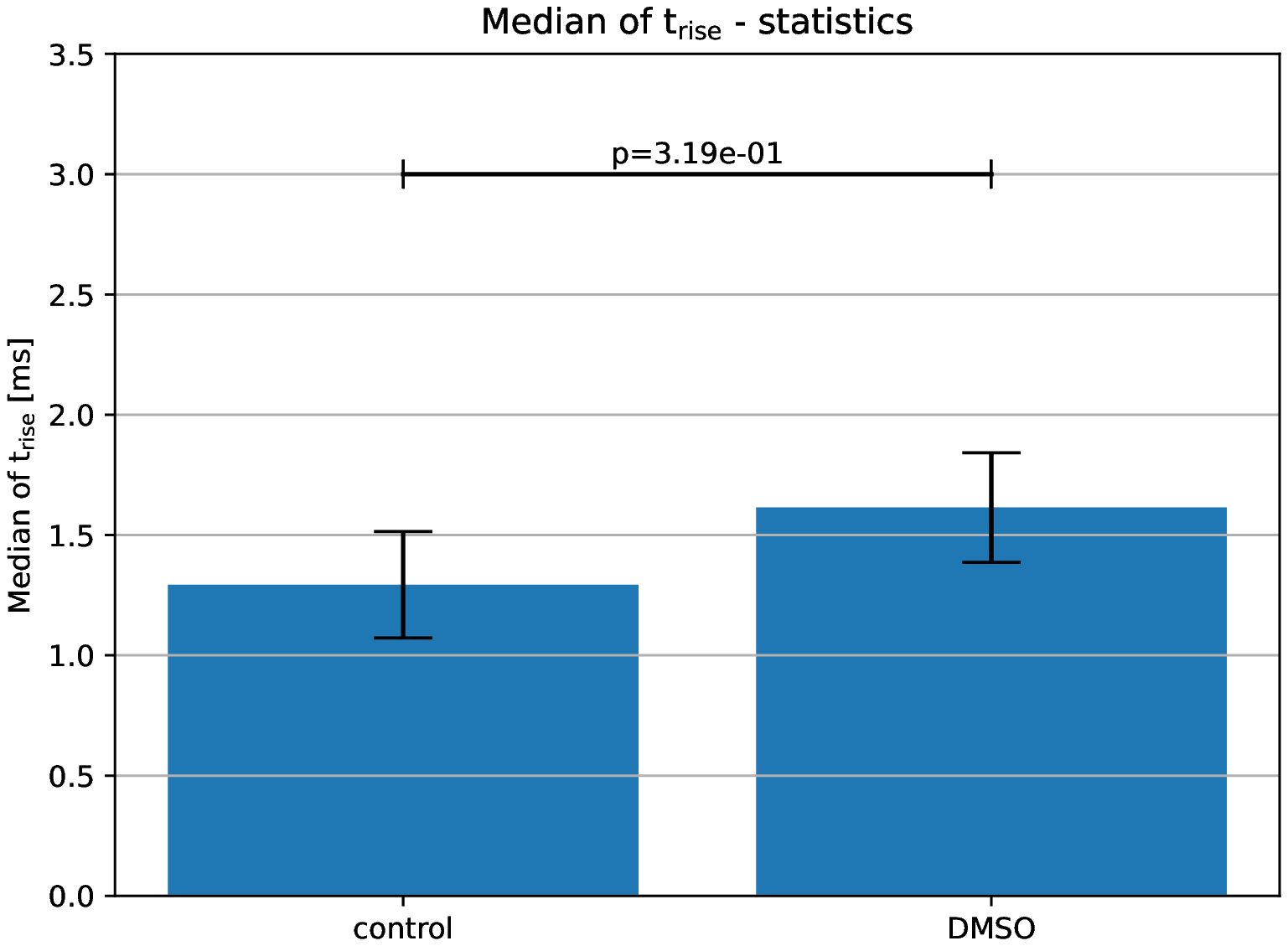}
\end{subfigure}
\begin{subfigure}{.4\linewidth}
\includegraphics[scale=0.4]{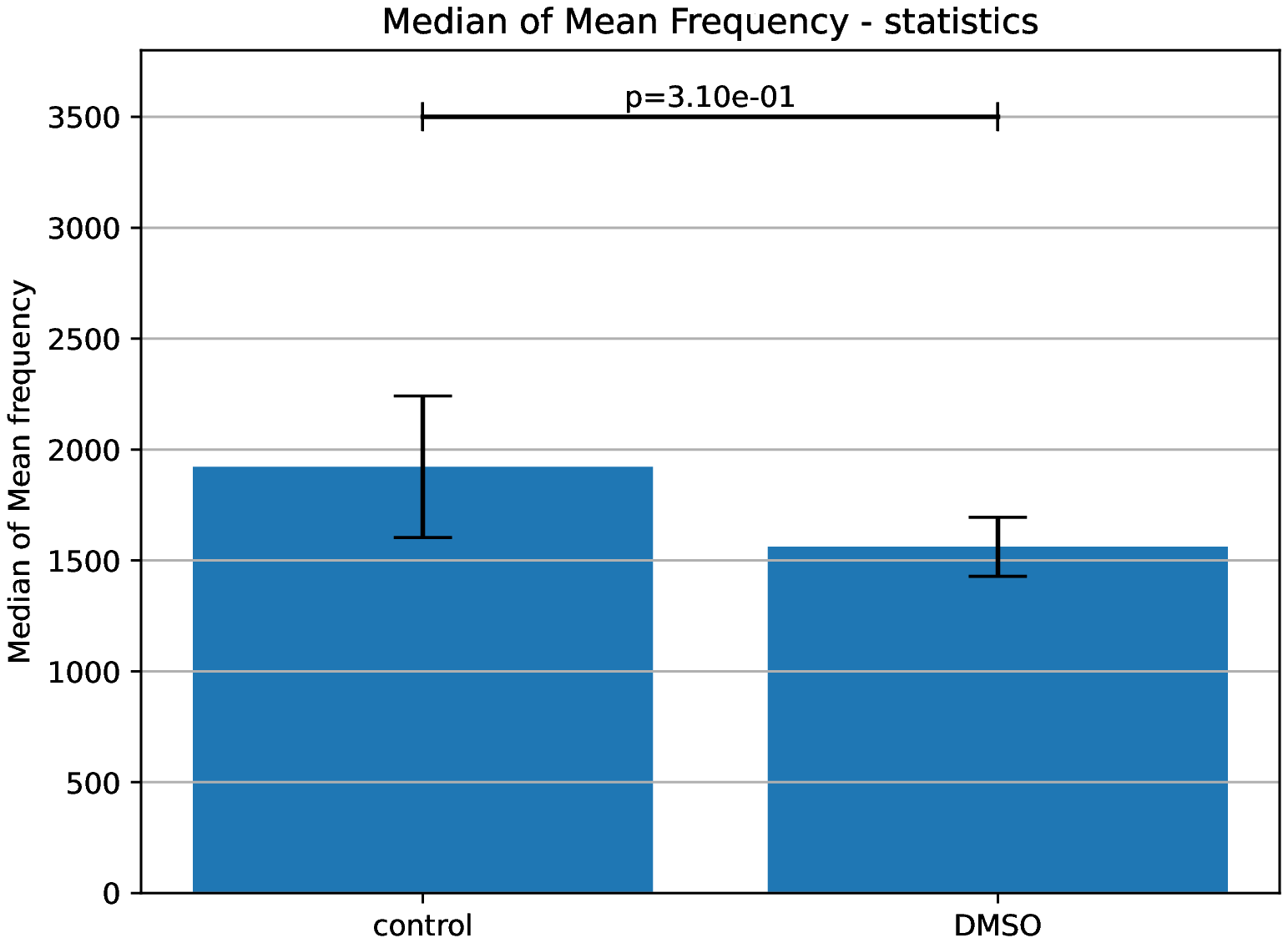}
\end{subfigure}
\caption{FFT on the DMSO dataset. Left: Spike characteristics for control and $0.6\%$ DMSO in the time domain shown by $t_\mathrm{rise}($25-100$\%)$. Right: Spike characteristics in the frequency domain shown by median of mean frequency, $f_\mathrm{mean}$. Standard error of mean of the median of $t_\mathrm{half}$ and the median of mean frequency are shown by the error bars.}
\label{fig:dmso_fmean_trise}

\begin{subfigure}{.4\linewidth}
\includegraphics[scale=0.4]{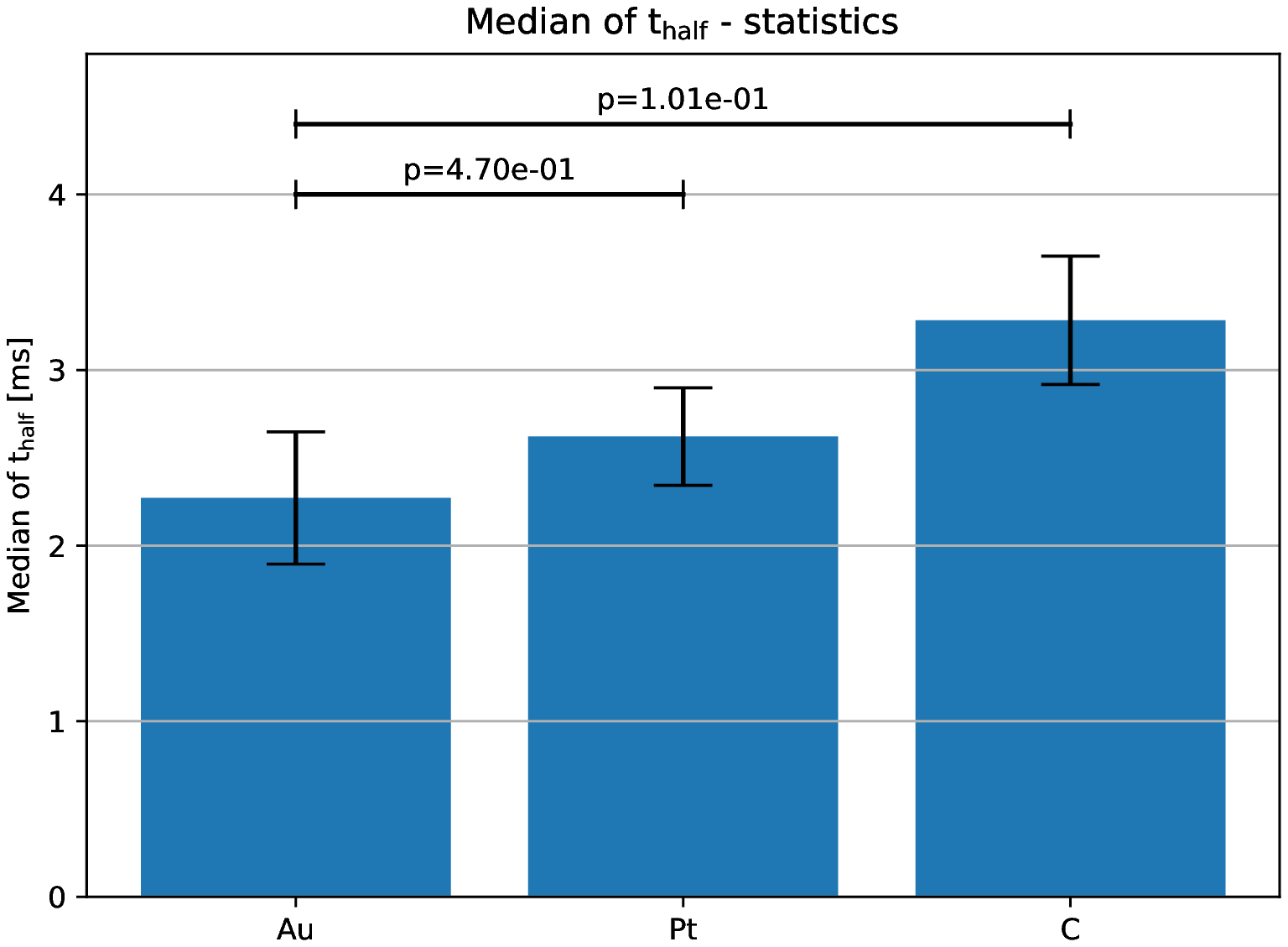}
\end{subfigure}
\begin{subfigure}{.4\linewidth}
\includegraphics[scale=0.4]{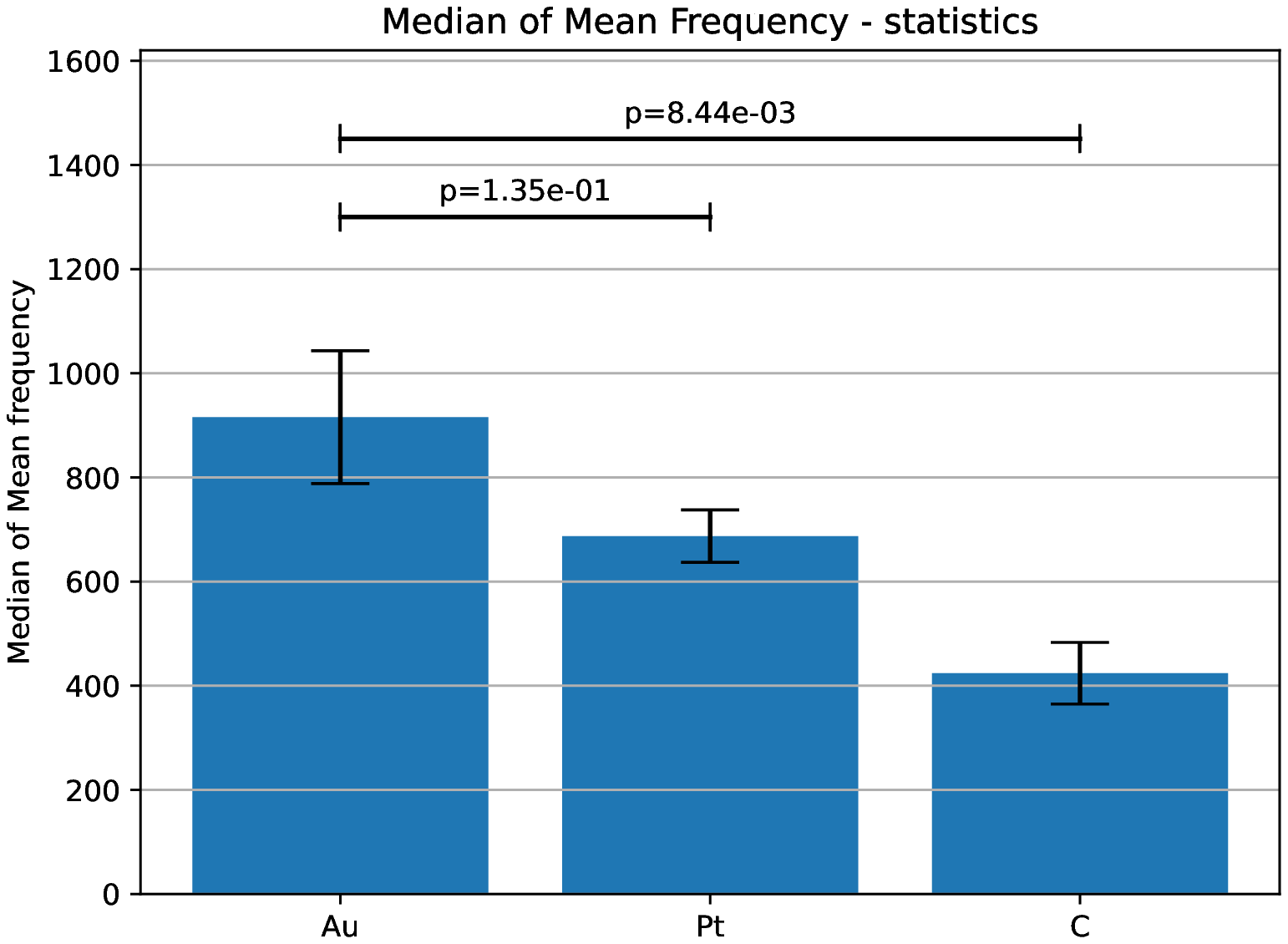}
\end{subfigure}
\caption{FFT on the electrodes dataset. Left: Spike characteristics due to different electrodes in the time domain shown by $t_\mathrm{half}$. Right: Spike characteristics in the frequency domain shown by median of mean frequency, $f_\mathrm{mean}$. Standard error of mean of the median of $t_\mathrm{half}$ and the median of mean frequency are shown by the error bars.}
\label{fig:electrodes}
\end{figure}
\subsection*{Observations on the Amperometry Datasets}
In the time domain, the mean of medians of spike width of the \hsd\ traces increases in the Hofmeister order, however with the exception of the nitrate ion (left panel of \fig\ \ref{fig:hof_fmean}). In the frequency domain, the mean frequency decreases along with the Hofmeister order, or increases exactly in the opposite order due to the inverse relationship between spike shape and mean frequency. The atypical ordering between the chlorate ion and nitrate ion is captured in the frequency domain as well (right panel of \fig\ \ref{fig:hof_fmean}). 

DMSO incubation influences only certain spike characteristics, specifically, it increases $t_\mathrm{rise}$. This is evident from the median of spike width in the time domain, where the control group shows a lower mean of the median value of $t_\mathrm{rise}$ compared to DMSO. In the frequency domain, DMSO shows lower mean frequency on average (\fig\ \ref{fig:dmso_fmean_trise}). Similar observations can be made on our third and final candidate, i.e. the electrodes dataset (\fig\ \ref{fig:electrodes}). 

The standard error of mean error bars per category, and the corresponding confidence intervals, reflect the effect of sample sizes ($5-10$ each category), as can been seen for example, in the electrodes dataset that has poor sample sizes. The datasets used here as candidates along with their respective sample sizes and other attributes are summarized in the supplementary section. 

This method is therefore amenable for statistical validation of any amperometry dataset in the frequency domain. However, it is important to note that (as can been seen from the electrodes dataset), the sample size and length of measurements play a key role as well. Too few samples or too short measurements per category may not be sufficient for statistical tests of frequency analysis, and may result in extremely low confidence intervals.

Although a workaround for this issue of large error bars might be removing outliers, this is highly discouraged since this would reduce the credibility of the observations made – in particular when the sample size is small. The procedure outlined herein is available as an open-source tool specifically dedicated to the analysis of amperometry signals in the frequency domain. The program implementation is also detailed in the supplementary section.
\section*{Conclusions}
\label{sec:concl}

In this work, we have outlined a method for spike-based frequency analysis of amperometry traces that also provides statistical validation of observations on spike characteristics in the time domain. To our knowledge, this is the first fully automated open-source tool available for analyzing amperometric spikes in the frequency domain. We have shown that the time-domain information could be retrieved from spike-based frequency analysis. The proposed method provides a more systematic way of analyzing amperometry data compared to IgorPro QuantaAnalysis which involves manual interventions with the GUI and on Excel. 

Further, we have outlined quite a diverse set of amperometry datasets that illustrates the relationship between spike shape and mean frequency. Although few steps of the frequency analysis method are user-dependent (such as data filtering and cut-off factor), the majority of our program is fully automated and has provided consistent results on different amperometric datasets. However, the frequency analysis implementation is a Python package that is not as mature as IgorPro QuantaAnalysis software which, for instance, has functionalities to handle the separation of overlapping spikes automatically. Further, the package does not have a GUI and interaction happens through a Command Line Interface. In addition, this method may not be suitable for datasets that have too few traces.

Further, though with Fourier Transform methods it is possible to evaluate all the frequencies in a signal, the time at which they occur cannot be determined since the signal is represented only in the frequency domain. This bottleneck may be overcome by using methods such as wavelet analysis which can represent a signal in the time and frequency domain at the same time\cite{Coifman92, Pavlov_2012}. Using discrete wavelet analysis on the detected spikes will give the time localization of the key frequency levels as well. 
\section*{Methods}
\label{sec:methods}

\subsection*{State-of-the-Art Amperometry Data Analysis Workflow}
As motivated in the introduction, amperometry traces are one-dimensional waveforms (or signals) that can also be treated in the frequency domain, thereby making frequency analysis an alternative to the traditional time-domain methods. The traditional analysis is usually done by using the mature spike-based amperometric data analysis software IgorPro QuantaAnalysis. In the state-of-the-art QuantaAnalysis workflow, amperometry traces are imported into the program, and traces are pre-processed using filters wherever necessary. 

Subsequently, the pre-processed data is used to identify spikes based on a pre-defined threshold. This is then followed by statistical analysis and visualization of over $20$ representative spike characteristics, including time parameters, such as $t_\mathrm{rise}$, $t_\mathrm{fall}$ and $t_\mathrm{\frac{1}{2}}$; current parameters, such as $I_\mathrm{max}$ and $I_\mathrm{foot}$; and charge parameters, such as $Q_\mathrm{foot}$ and $Q_\mathrm{spike}$ (as shown in top sequence of workflow in \fig\ \ref{fig:workflow}(A) and illustrated in \fig\ \ref{fig:spike}). In addition, the spike characteristics of all spikes in an amperometry trace are written out by the QuantaAnalysis program to a table similar to that shown in \fig\ \ref{fig:workflow}(B).

\subsection*{Frequency Analysis Workflow}
\begin{figure}[!htbp]
\centering
\textbf{(A)}
\begin{tikzpicture}[node distance = 3.0cm, auto]
  \tikzstyle{decision} = [diamond, draw, fill=blue!20,
    text width=5em, text badly centered, node distance=1cm, inner sep=0pt]
  \tikzstyle{block} = [rectangle, draw,
    text width=5.8em, text centered, rounded corners, minimum height=2em]
  \tikzstyle{line} = [draw, -latex']  
  \tikzstyle{cloud} = [draw, ellipse,node distance=3cm,
    minimum height=1em, text width=5em]
    
    \node[block] (data) {Import amperometry dataset};
    \node[block, right of=data] (pre) {Pre-processing};
    \node[block, above right of=pre, , fill=gray!20] (igor) {IgorPro spike analysis};
    \node[block, right of=igor, , fill=gray!20] (char) {Collect spike characteristics};
    \node[block, right of=char, , fill=gray!20] (vis) {Visualize median of $t_\mathrm{half}, t_\mathrm{rise}, t_\mathrm{fall}$ statistics};

    \node[block, below right of=pre, fill=blue!20] (extract) {Identify and extract spikes};
    \node[block, right of=extract, fill=blue!20] (fft) {Run FFT on every spike};
    \node[block, right of=fft, fill=blue!20] (fmean) {Evaluate mean frequencies of all spikes};
    \node[block, right of=fmean, fill=blue!20] (fmed) {Visualize median of mean frequency statistics
    };
    
    \path [line, thick] (pre) --node [text width=6cm,midway,above] {\footnotesize{State-of-the-art}} (igor);
    \path [line, thick] (pre) --node [text width=6cm,midway,below] {\footnotesize{Frequency Analysis}} (extract);
    \path [line, thick] (data) -- (pre);
    \path [line, thick] (igor) -- (char);
    \path [line, thick] (char) -- (vis);
    \path [line, thick] (extract) -- (fft);
    \path [line, thick] (fft) -- (fmean);
    \path [line, thick] (fmean) -- (fmed);
\end{tikzpicture}
\textbf{(B)}
\begin{tabular}{ |c|c|c|c|c|c|c| } 
 \hline
  & $t_\mathrm{1/2}$ [ms] & $t_\mathrm{fall}$ [ms] &$t_\mathrm{max}$[s] & $I_\mathrm{foot}$ (pA) & $t_\mathrm{foot}$[ms] &$t_\mathrm{rise}$[ms] \\ 
  \hline
   $\mathrm{spike}_1$ & \ldots & \ldots &\ldots & \ldots & \ldots& \ldots\\ 
  $\mathrm{spike}_2$ & \ldots &\ldots  &\ldots & \ldots &\ldots & \ldots\\ 
  \vdots & \ldots &\ldots  &\ldots & \ldots & \ldots& \ldots\\ 
  $\mathrm{spike}_{n_1}$ & \ldots & \ldots & \ldots& \ldots & \ldots& \ldots\\ 
 \hline
\end{tabular}\\ \vspace{0.2cm}
\textbf{(C)}
\begin{tabular}{|c|c|c|c|}
 \hline
 trace ID & spike ID& $f_\mathrm{mean}$ & median($f_\mathrm{mean}$)\\
 \hline
 & $\mathrm{spike}_1$ & \ldots & \\
1& $\mathrm{spike}_2$ & \ldots & \ldots\\
 & \vdots & \ldots & \\
 & $\mathrm{spike}_{n_1}$ & \ldots & \\
\hline
 & $\mathrm{spike}_1$ & \ldots & \\
2& $\mathrm{spike}_2$ & \ldots & \ldots\\
& \vdots & \ldots & \\
& $\mathrm{spike}_{n_2}$ & \ldots & \\
\hline
\end{tabular}
\caption[Workflow Pipeine]{Analysis of amperometry traces: (A) a comparison between the workflow adopted by the \enquote{state-of-the-art} method, which uses QuantaAnalysis to analyze time-domain spike parameters of the trace, and the proposed \enquote{frequency analysis} method, which analyzes the spikes in the frequency domain. (B) shows an overview of the output time, current and charge parameters in the time domain. (C) shows an overview of the output parameters in the frequency domain, mean frequencies and their median within each trace. The frequency analysis method provides a statistical validation for the state-of-the-art.}
\label{fig:workflow}
\end{figure}
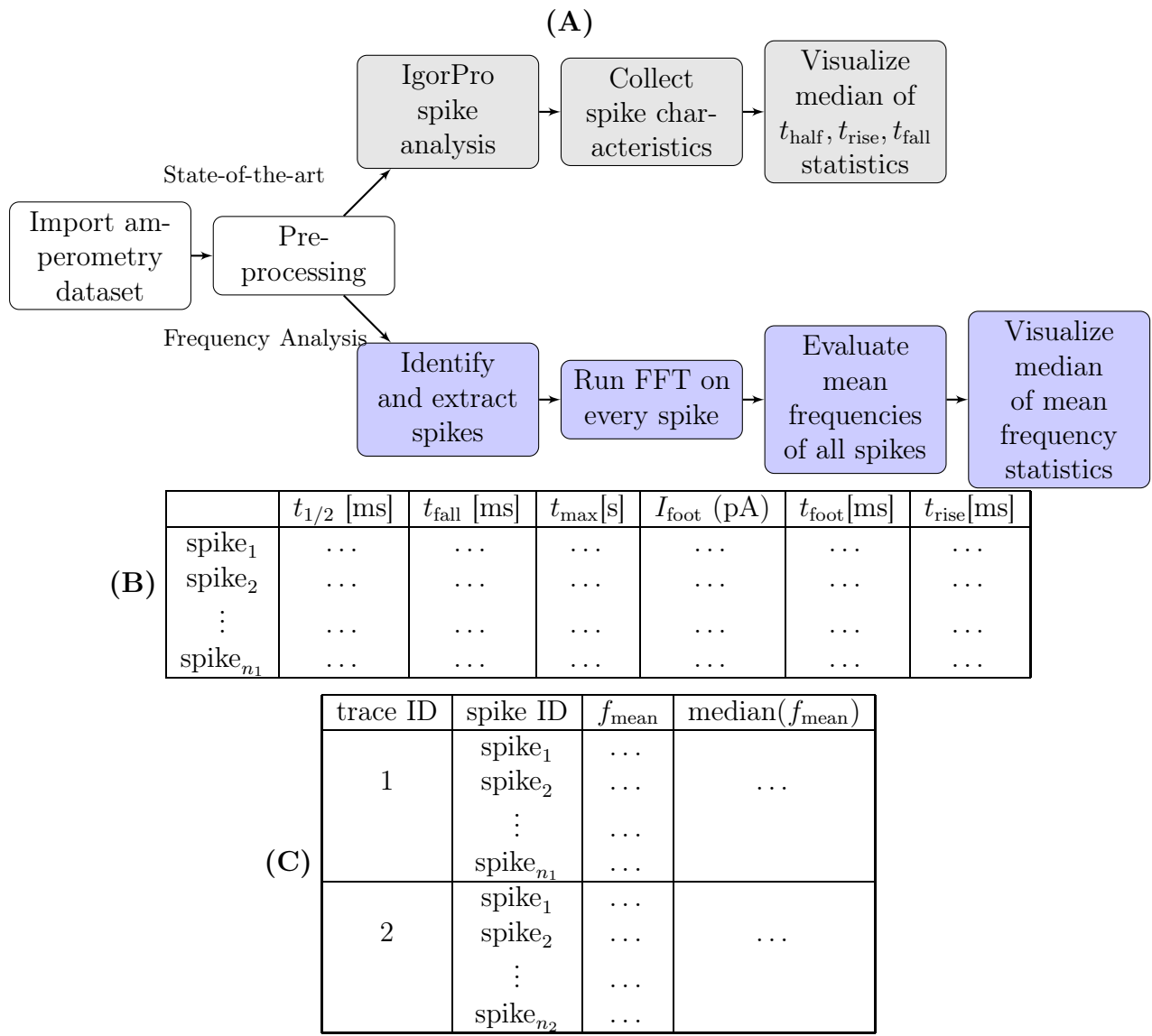

In the proposed, \enquote{frequency analysis} workflow shown in the bottom sequence of \fig\ \ref{fig:workflow}(A), similar pre-processing is done by removing unstable data caused by poor device alignment, followed by low-pass filtering for noise removal. Few signals could be exceptions to automated pre-processing and may require manual intervention. This may include cases where there are artifacts (\eg\ oscillations or \enquote{jumps}) at the beginning of the trace, or spike clusters occurring from time to time, and the presence of noise arising due to the type of electrodes used in the experiment. Signal \enquote{jumps} at the beginning of the signal were for instance observed in the electrodes dataset, perhaps due to poor alignment of the measuring instruments. 

This may lead to unrealistic spikes that may crash the program. In addition, we also observed spike clusters occasionally occurring in the datasets, wherein hundreds or even thousands of spikes occur sequentially. Such clusters can cause significant bias when evaluating the spike statistics. These anomalies were observed with varying probabilities of occurence for all datasets we analyzed, usually caused by poor device alignment.  

In these exceptional cases, we recommend manual handling of these traces; for instance by ignoring the starting segment of the traces that contain artifacts or by ignoring spikes identified as belonging to a cluster. Once this is done, the standard deviation of the baseline, $\sigma_{base}$ and local maxima are determined. Then each local maximum is analyzed and compared with a pre-defined threshold factor $\times$ $\sigma_{base}$. Using this threshold, spikes are identified and extracted, followed by the application of Fast Fourier Transform (FFT) on each spike (the FFT method is discussed in more detail in the latter part of this section). 

Using FFT, the mean frequency of each spike can be calculated in the frequency domain. Subsequently, the means of median of the mean frequencies of all spikes (and corresponding standard deviation) can be used for statistical analysis of the traces (as shown in \fig\ \ref{fig:workflow}(C)). This pipeline in the frequency domain, outlined above, is implemented as an end-to-end Python package that takes in raw amperometry trace and outputs the mean frequencies and their statistics. A more detailed treatment of the package implementation is outlined in the supplementary section.

\subsection*{Fourier Transform}
Amperometry traces are signals that are sampled discretely in time, hence we require a Discrete Fourier Transform (or DFT) for frequency analysis. The DFT is a commonly used method in signal processing to convert a signal in the time domain into a frequency spectrum, enabling us to characterize the mean or dominant frequency properties of the signal. Mathematically, a Fourier Transform breaks down a non-linear function into a linear superposition of simple basis functions, such as sines and cosines, whose summation reconstructs the original signal. 

The basis function herein must form an orthonormal base to ensure linear independence, which are the complex exponentials in the Fourier Transform. In terms of computational cost, a DFT requires $O(N^2)$ operations, which gets quite expensive for a large number of sampling points. The Fast Fourier Transform method numerically divides the DFT into smaller DFTs thereby bringing down the number of operations to complexity of $O(N \log N)$.

Consider a periodic sinusoidal signal, $f(t) = A \sin(2\pi \omega t)$, where $A$ is the amplitude of the signal and $\omega$ is the angular frequency of the signal. A Fourier Transform of this signal is exactly supported at $\omega$ and $-\omega$. For such a signal, if the waveform oscillates rapidly as a function of time, it is referred to as a high frequency signal (such as $s_1$ in \fig\ \ref{fig:hypothesis}), else if it varies slowly then it is referred to as low frequency signal (such as $s_2$ in \fig\ \ref{fig:hypothesis}). 

\begin{figure}[!htbp]
\centering
\includegraphics[scale=0.6]{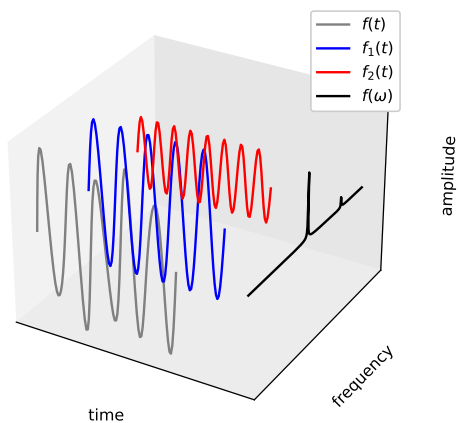}
\caption{Illustration above represents the idea of a Fourier Transform: Consider a signal, $f(t) = \sin{(10\pi t)} + 0.2\cos{(16\pi t)}$ in the time-amplitude space(represented by the gray curve). This signal can be decomposed into $f_1(t) = \sin{(10\pi t)}$ and $f_2(t) = 0.2\cos{(16\pi t)}$ in the time-amplitude space (represented by the blue and red curves respectively). Fourier Transform of the signal, $f(\omega)$ shows these two peak frequencies respectively in the frequency-amplitude space (represented by the black curve).}
\label{fig:fft}
\end{figure}

In the case of linear superposition of multiple signals, such as the one illustrated in \fig\ \ref{fig:fft}, Fourier Transform can reveal the constituent trigonometric functions at different frequencies decomposed from these signals. The discrete Fourier transform transforms a sequence of $N$ complex numbers, $x_n$ into another sequence of numbers, $X_k$, which is given by,

\begin{equation*}
    x_n  = \frac{1}{N} \sum_{k=0}^{N-1} X_k\exp^{\frac{2\pi j k n}{N}}
\end{equation*}
and
\begin{equation*}
X_k = \sum_{n=0}^{N-1}x_n \exp^{\frac{-2\pi jkn}{N}}
\end{equation*}

where $N$ is the number of time samples we have from the trace; $n$ is the current sample we are considering; $\frac{2 \pi k}{N}$ is the current frequency ($0$ to $(N-1)$ \hz) and $X_k$ is the amount of frequency $k$ in the signal (constituted by amplitude and phase). A non-oscillating shift only changes the coefficient $X_0$. Therefore the baseline shift can be neglected by ignoring the $0$ \hz\ component in FFT.  Please note that usually a factor of $\frac{1}{N}$ is used in the inverse transform from frequency to the time domain. Alternatively, one could apply the factor $\frac{1}{\sqrt{N}}$ to both DFT and inverse DFT. 

\subsection*{Main and Mean Frequency}

For our analysis of amperometry traces, the frequency is in the range of $[0, f_\mathrm{sampling}]$. The experimental data were recorded with a sampling frequency, $f_\mathrm{sampling} = 10$ \khz, therefore the Nyquist frequency is one-half of the sampling frequency \ie\ $5$ \khz. When considering only the frequency components below the Nyquist frequency, the resulting discrete-time signal can be exactly reconstructed without distortion (known as aliasing). 

The main frequency that comes out of the analysis is the one with the largest amplitude, whereas the mean frequency is calculated as the energy averaged frequency, $f_{\mathrm{mean}}=\frac{\sum k*\mathrm{amp(X_k)^2}}{\sum \mathrm{amp(X_k)^2}}$. Note that analyzing the full-time series (including baseline) would show no characteristic frequencies apart from a peak at around $0$ \hz\ (refer \fig\ \ref{fig:fft_whole} of the supplementary section). However, doing a spike-based frequency analysis gives true insight into the mean frequency range of the traces. 

Once this implementation for frequency analysis is established, it is straightforward to observe the reflection of spike shape (thin or wide) on the mean frequency. The relationship between spike shape and mean frequency has been motivated by way of simulated signals in the conclusions section. It can also be shown analytically using high- and low-frequency triangular signals in the time domain as objects for Fourier Transform (detailed in the supplementary section). 
\section*{Funding}

This work was supported by the JPND Neuronode grant No. 01ED1802 for the computational work; and  the MSCA grant funding from the European Union's Horizon 2020 research and innovation program under the Marie Skłodowska-Curie Grant Agreement No.793324 for the experimental work.
\begin{acknowledgement}
The authors thank Mohaddeseh Aref, Zahra Taleat, Alex S. Lima, Elias Ranjbari, Johan Dunevall for their data contributions.
\end{acknowledgement}
\begin{suppinfo}
\label{sec:supp}
\subsection*{Analytical Verification}
\label{subsec:analytical}

Calculating FFT over the entire time series (including baseline) would show no characteristic frequencies apart from a peak at around $0$ \hz\ since the slowing varying baseline dominates the signal, as an example, see \fig\ \ref{fig:fft_whole}. 

\begin{figure}[!htbp]
    \centering
     \includegraphics[scale=0.5]{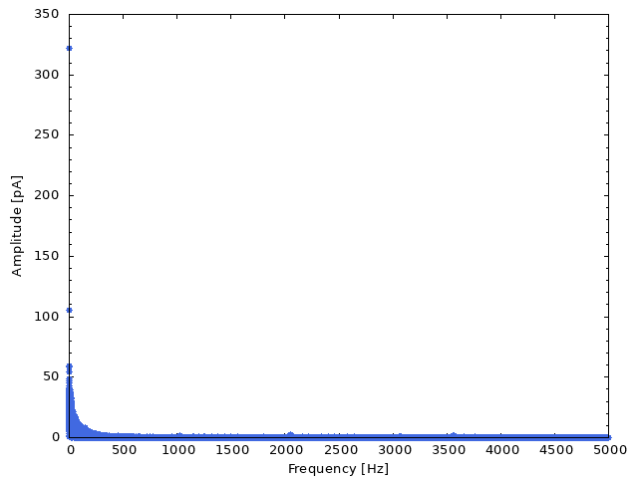}
    \caption{Fast Fourier Transform of a $\mathrm{Br}^-$ time series. All other data also show a peak frequency component near $0$ Hz.}
    \label{fig:fft_whole}
\end{figure}

Therefore, the FFT should be applied on a spike-by-spike basis to extract the spike-specific mean frequencies. As motivated in the previous sections, the frequency analysis method is introduced based on the hypothesis that a relation between spike shape and the mean frequency exists. We here show the proof of this hypothesis for two simple waveforms, \ie\ a sinusoidal waveform and a symmetric triangle waveform.
\subsubsection*{Sinusoidal Waveform}
\label{subsubsec:sine}

For a given signal, $y(t)$, the signal can be transformed to the frequency domain, $\ft(y(t)) = Y(\s)$ using the Fourier Transform. Next, to compute the mean frequency, $f_\mathrm{mean}$, one needs to calculate the ratio of the first moment of $|Y(\s)|^2$ and the expectation of $|Y(\s)|^2$, which is given by,

\begin{equation}
\fm(\omega) = \frac{\int_{0}^{\infty}|Y(\s)|^2~\s~d\s}{\int_{0}^{\infty} |Y(\s)|^2~d\s}
\label{eq:fmean}
\end{equation}

where the $|Y(\s)|$ is given by,

\begin{equation}
|Y(\s)| = \sqrt{\mathrm{Re}(Y(\s))^2 + \mathrm{Im}(Y(\s))^2}.
\label{eq:ys}
\end{equation}

\begin{figure}[!htbp]
\includegraphics[scale=0.8]{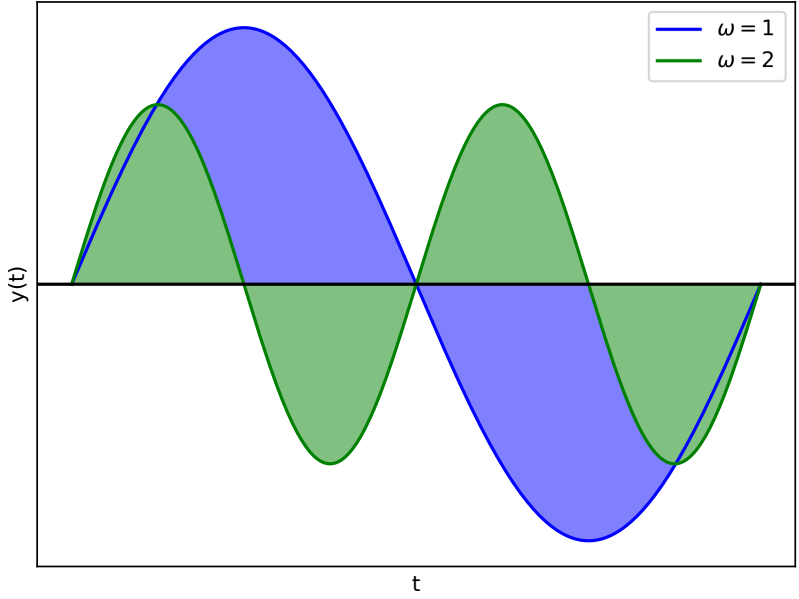}
\caption{Illustration of sine waves with different frequencies in time domain.}
\label{fig:sine}
\end{figure}

Let us consider an example of the waveform, $\sin(\omega t)$ as illustrated in \fig\ \ref{fig:sine},

\begin{equation}
\begin{aligned}
Y(\s) &= \int_{0}^{2\pi} y(t)~e^{-2\pi it \s}~dt\\
&= \int_{0}^{2\pi} \sin(\omega t)~e^{-2\pi it \s}~dt\\
&= \frac{1}{2i}\Big[ \delta \big( \s - \frac{\omega}{2\pi}\big) - \delta\big( \s + \frac{\omega}{2\pi} \big)\Big]\\
\end{aligned}
\label{eq:yssine1}
\end{equation}

\begin{equation}
\begin{aligned}
|Y(\s)| &= \sqrt{\mathrm{Re}(Y(\s))^2 + \mathrm{Im}(Y(\s))^2}\\
&= \frac{1}{2} \Big[ \delta \big( \s - \frac{\omega}{2\pi}\big) - \delta\big( \s + \frac{\omega}{2\pi} \big)\Big]\\
\end{aligned}
\label{eq:yssine2}
\end{equation}

\begin{equation}
|Y(\s)| = \delta \big( \s - \frac{\omega}{2\pi}\big)
\end{equation}

Mean frequency can then be calculated as,

\begin{equation}
\begin{aligned}
\fm(\omega) &= \frac{\int_{0}^{\infty}|Y(\s)|^2~\s~d\s}{\int_{0}^{\infty} |Y(\s)|^2~d\s}\\
&= \frac{\int_{0}^{\infty} \delta(\s - \frac{\omega}{2\pi})~\s~d\s}{\int_{0}^{\infty} \delta(\s- \frac{\omega}{2\pi} \big)~d\s}\\
\fm(\omega) &= \frac{\omega}{2\pi}\\
\end{aligned}
\label{eq:fmeansine}
\end{equation}

As expected, we get a formulation for the mean frequency as being directly proportional to $\omega$. 
\subsubsection*{Triangle Waveform}
\label{subsubsec:tri}

Now we can consider a waveform that represents a symmetric amperometry spike with a linear rise and fall. A triangular wave or triangle wave is a periodic, piecewise linear, continuous real function (also referred to as the hat function) such as that illustrated in \fig\ \ref{fig:hat}. 

\[
    y(t)= 
\begin{cases}
    1- \frac{|t|}{a}, & |t| \leq a\\
    0,              & \text{otherwise}
\end{cases}
\]

\begin{figure}[!htbp]
\includegraphics[scale=0.8]{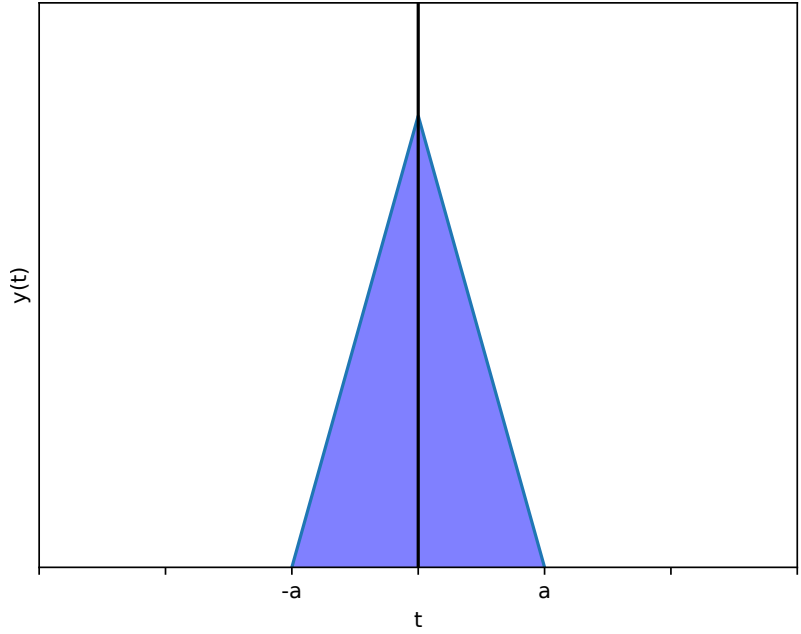}
\caption{Illustration of a symmetric hat function in the time domain with base length $2a$.}
\label{fig:hat}
\end{figure}

Following the steps outlined in the previous subsection, we can find the relationship between mean frequency and length of the base of triangle as,

\begin{equation}
\begin{aligned}
\ft(y(t)) &= \int_{-\infty}^{\infty} e^{-2\pi i\s t}dt \\
&= \int_{-a}^{0} \Big(1 + \frac{t}{a} \Big)~e^{-2\pi i \s t}~dt + \int_{0}^{a} \Big(1 - \frac{t}{a} \Big)~e^{-2\pi i \s t}~dt\\
\ft(y(t) &= a^2 \sinc^2(a~\s)\\
\end{aligned}
\label{eq:ystri1}
\end{equation}

\begin{equation}
|y(\s)| = 2 a \sinc^2(a\s)
\label{eq:ystri2}
\end{equation}

Mean frequency,

\begin{equation}
\begin{aligned}
\fm &= \frac{\int_{0}^{\infty}|Y(\s)|^2~\s~d\s}{\int_{0}^{\infty} |Y(\s)|^2~d\s}\\
&= \frac{3}{a} \ln 2 \pi\\
\end{aligned}
\label{eq:fmeantri}
\end{equation}

Therefore, we can see that larger the $a$ then wider is the spike, and hence lower is its mean frequency, as one might expect. 
\subsection*{Code}
\label{subsec:code}

The \texttt{frequency-analysis} package is implemented in Python3.2.0 and the most recent version can be downloaded from: \href{https://github.com/krishnanj/SSFAAT}{https://github.com/krishnanj/SSFAAT}. The package consists of four modules explained in \fig\ \ref{fig:package}.

\begin{figure}[!htbp]
\centering
\begin{tikzpicture}[node distance = 3.5cm, auto]
  \tikzstyle{decision} = [diamond, draw, fill=blue!20,
    text width=5em, text badly centered, node distance=1cm, inner sep=0pt]
  \tikzstyle{block} = [rectangle, draw,
    text width=5.8em, text centered, rounded corners, minimum height=2em]
  \tikzstyle{line} = [draw, -latex']  
  \tikzstyle{cloud} = [draw, ellipse,node distance=3cm,
    minimum height=1em, text width=5em]
    
    \node[] at (0,0) (raw) {visualization};
    \node[right of=raw] (auto) {automator};
    \node[right of=auto] (main) {main};
    \node[right of=main] (utils) {utils};

    \path [line, thick] (raw) -- (auto);
    \path [line, thick] (auto) -- (main);
    \path [line, thick] (utils) -- (main);

\end{tikzpicture}
\caption{Structure of the spike-based frequency analysis Package. All programs fetch parameters set by the use in the \texttt{params} module.}
\label{fig:package}
\end{figure}
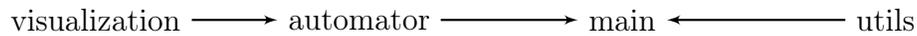 

The \texttt{main} module uses the helper functions defined in \texttt{utils} to do spike-by-spike mean frequency evaluation for each category. Further, it shows a visualization of the distribution of mean frequencies. The user could change these parameters to analyze their data. The \texttt{raw\_data\_visualization} module shows visualizations of the input raw data.

The \texttt{params} module defines the global parameters, such as the paths to raw data, interim data, sampling frequency and output path as outlined below in \fig\ \ref{fig:params}. 

\begin{figure}
\begin{Verbatim}[frame=single]
// General parameters
num_categories  # number of categories or labels
category_list 	# list of strings with label ids
path_raw_category    # path to raw data of each category
target_path_category # write path of the excel sheet that contain 
                       spike features for each category

// FFT parameters
f_samp 			# sampling frequency

// Peak detection parameters
thrs  			# threshold for peak detection

// Other
exclusion_list  # list of file names with dubious data
\end{Verbatim}
\caption{Overview of parameters that requires user intervention in the \texttt{params} module.}
\label{fig:params}
\end{figure}

\clearpage
The \texttt{automator} module identifies spikes based on a pre-defined threshold and extracts spikes. A summary of flow of the program in the form of a pseudocode is outlined as follows:

\begin{algorithm}[!htbp]
  \caption{The \texttt{automator} module executes the method \texttt{peak\_detector\_automated} using user-inputs including list of label categories, path to raw data, path to output files and peak threshold set in \texttt{params} module. }
  \label{code:automator}
  \begin{algorithmic}
    \Procedure{peak\_detector\_automated}{category\_list, raw\_path\_list, target\_path\_list, peak\_threshold}
    \For{each category in category\_list}
    	\State raw\_path\_category = raw\_path\_list[category]
    	\For{every file in raw\_path\_category}
    		\If{data file not in exclusion\_list}
    			\State load raw data
    			\State find standard deviation of baseline from first $30$ data points
    			\State calculate peaks based on peak\_threshold; get peak properties and store
    			\If{peaks exist}
    				\State calculate normalized cutoff position and apply time series filtering
    				\State extract t\_max, I\_start, I\_end of the spikes
    				\State determine t\_half, t\_rise and t\_fall window
    			\EndIf
    			\State{Write all extract spike parameters to file in target\_path\_list and store}
    		\EndIf
    	\EndFor
    \EndFor
    \State return \texttt{output\_file}
    \EndProcedure
  \end{algorithmic}
\end{algorithm}

\clearpage
The \texttt{utils} module contains the helper functions for the FFT analysis and performs the following functions: \begin{enumerate*} \item given the raw time series and sampling frequency, \texttt{meanfreq} evaluates the mean frequency \item for each time series together with the pre-computed characteristic spike features of a given category, \texttt{analyze\_meanfreq} evaluates the mean frequency of each individual spike event w.r.t. a given interested window (e.g. $\mathrm{t_{half}}$ window)\item for a given simulated signal, \texttt{analyze\_meanfreq\_artificial} evaluates the mean frequency of each spike w.r.t. a given interested window and returns the median of mean frequencies \item \texttt{resample} function resamples each spike from its original length to a new length scaled by a factor \item \texttt{create\_spike\_train} generates an artificial spike train of given length and sampling frequency, wherein the spike geometry is specified in the supplementary section on artificial data generation. \end{enumerate*} In the following, we will focus on the core method in this module, \texttt{analyze\_meanfreq}:

\begin{algorithm}[!htbp]
  \caption{The \texttt{utils} module executes the method \texttt{analyze\_meanfreq} using user-inputs including the path to raw data directory and path to excel sheet.}
  \label{code:utils}
  \begin{algorithmic}
    \Procedure{analyze\_meanfreq}{path\_raw, path\_excel}
    \State initialize an empty 2-D array to write out mean\_freq
    \For{every file in path\_raw}
    	\State read in time series from file
    	\State read in spike parameters from path\_excel 
    	\State read in start and end time of the interested window for each spike, t\_start \\\hspace{1cm} and t\_end
    	\State convert time parameters into array index, index\_start and index\_end
    	\State determine window\_data for each spike
    	\For{each extracted spike}
    		\State execute \texttt{meanfreq(window\_data, f\_samp)} method \\\hspace{2cm} and store into mean\_freq[file\_index][spike\_index]
    	\EndFor
    \EndFor
    \State return \texttt{mean\_freq}
    \EndProcedure
  \end{algorithmic}
\end{algorithm}



\clearpage
\subsection*{On raw data}
\label{subsec:raw}

A summary of the attributes of the candidate datasets used in our analysis is given in Table \ref{tab:meta}. A short summary of the procedure adopted to generate these datasets is summarized from \cite{He, Majdi2017} in the following subsections:

\begin{table}[!htbp]
\centering
\begin{tabular}{ |c|p{50mm}|p{30mm}|p{30mm}| } 
 \hline
 Attribute & Hofmeister Series & Electrodes & DMSO \\ 
 \hline \hline
 Method & SCA & VIEC & IVIEC \\ 
 \hline
 Cell/ vesicle type & Adrenal chromaffin cells & chromaffin vesicles & chromaffin cells\\ \hline
 Conditions & $K^+$ and anions stimulation & Dipping Method & Control stimulated with $\mathrm{Ba}^{2+}$, additionally incubated with $0.6\%$ DMSO \\ \hline
 Categories & Hofmeister Ions  & Au, Pt, C & Control, DMSO\\ \hline
 Length (sec) & $40 - 126$ & $105 - 1459$ & $795 - 1544$\\ \hline
 Sampling Frequency (kHz) & $10$ & $10$ & $10$\\ \hline
 \# Samples & $158$ & $22$ & $21$\\ 
 \hline
 \# Samples per category & 
 \begin{itemize} 
 	\item $\mathrm{Br}^-$ - $26$ 
 	\item $\mathrm{Cl}^-$ - $29$ 
 	\item $\mathrm{ClO}_4^-$ - $30$ 
 	\item $\mathrm{NO}_3^-$ - $31$ 
 	\item $\mathrm{SCN}^-$ - $27$ 
 	\end{itemize} 
 & 
\begin{itemize} 
 	\item C - $4$ 
 	\item Pt - $7$ 
 	\item Au - $7$ 
\end{itemize} 
 & 
\begin{itemize} 
 	\item Control - $16$ 
 	\item DMSO - $21$ 
\end{itemize} \\
 \hline
\end{tabular}
\caption{Summary of attributes of candidate datasets}
\label{tab:meta}
\end{table}

\subsubsection*{Hofmeister Series Dataset}
Bovine adrenal glands were obtained from a local slaughterhouse and the cells were kept at $37^\circ C$ in isotonic solution during the whole experimental process. Electrochemical recordings from single chromaffin cells were performed on an inverted microscope in a Faraday cage. The working electrode was held at $+ 700$ mV versus an Ag/AgCl reference electrode and the output was filtered at $2.1$ kHz and digitized at $5$ kHz. For single-cell exocytosis, the micro-disk electrode was moved slowly by a  patch-clamp micromanipulator to place it on the membrane of a chromaffin cell without causing any damage to the surface. Ten seconds after the start of recording, $30$ mM $K^+$ stimulating solution in a glass micropipette was injected into the surrounding of the chromaffin cells with a single $30$-s injection pulse.

\begin{figure}[!htbp]
    \centering
    \begin{subfigure}[b]{0.3\textwidth}
        \centering
        \includegraphics[scale=0.3]{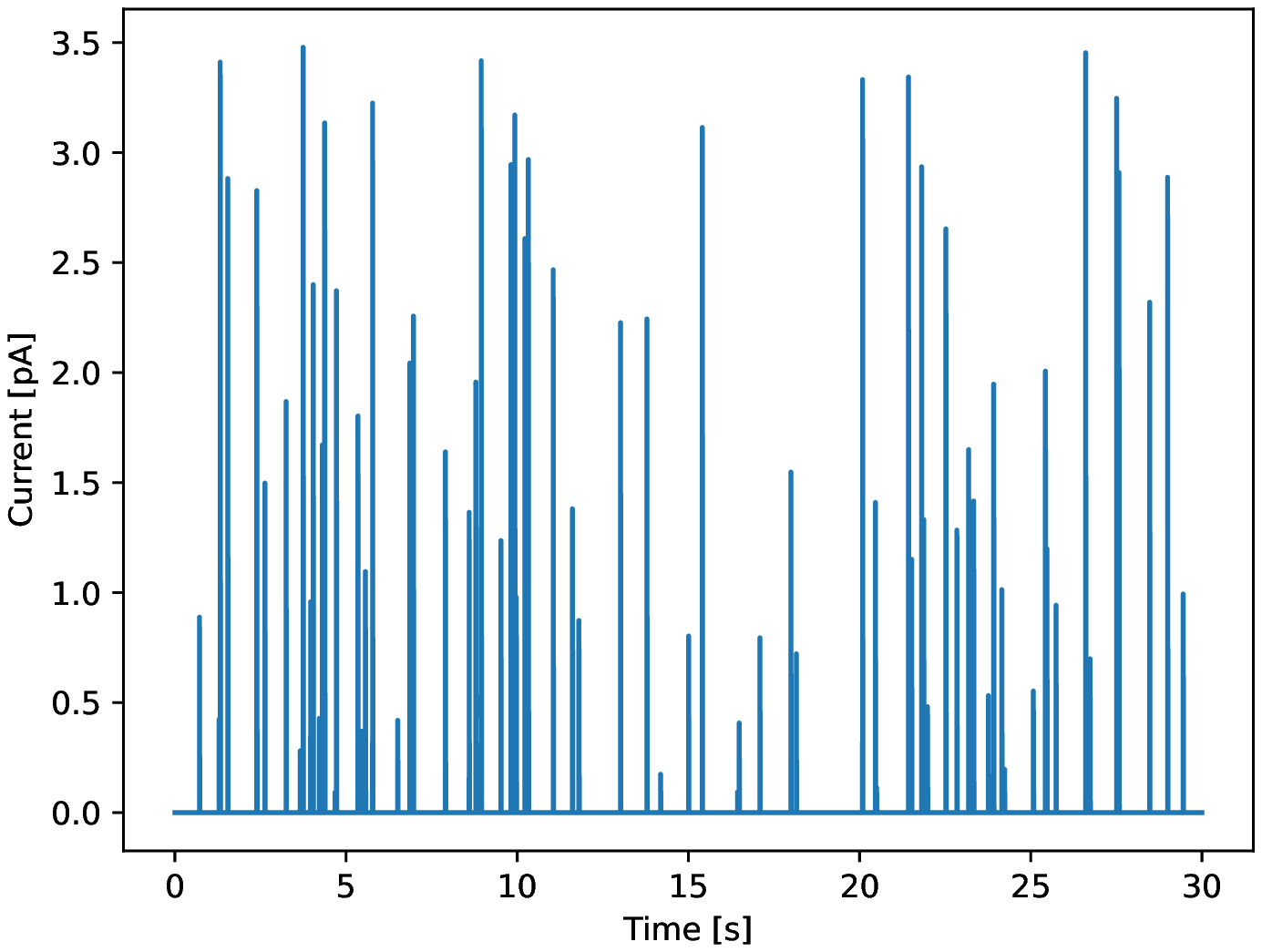}
        \caption[]%
        {{\small Spike trains with Br$^-$ stimulation}}    
        \label{fig:br}
    \end{subfigure}
    \begin{subfigure}[b]{0.3\textwidth}
        \centering
		\includegraphics[scale=0.3]{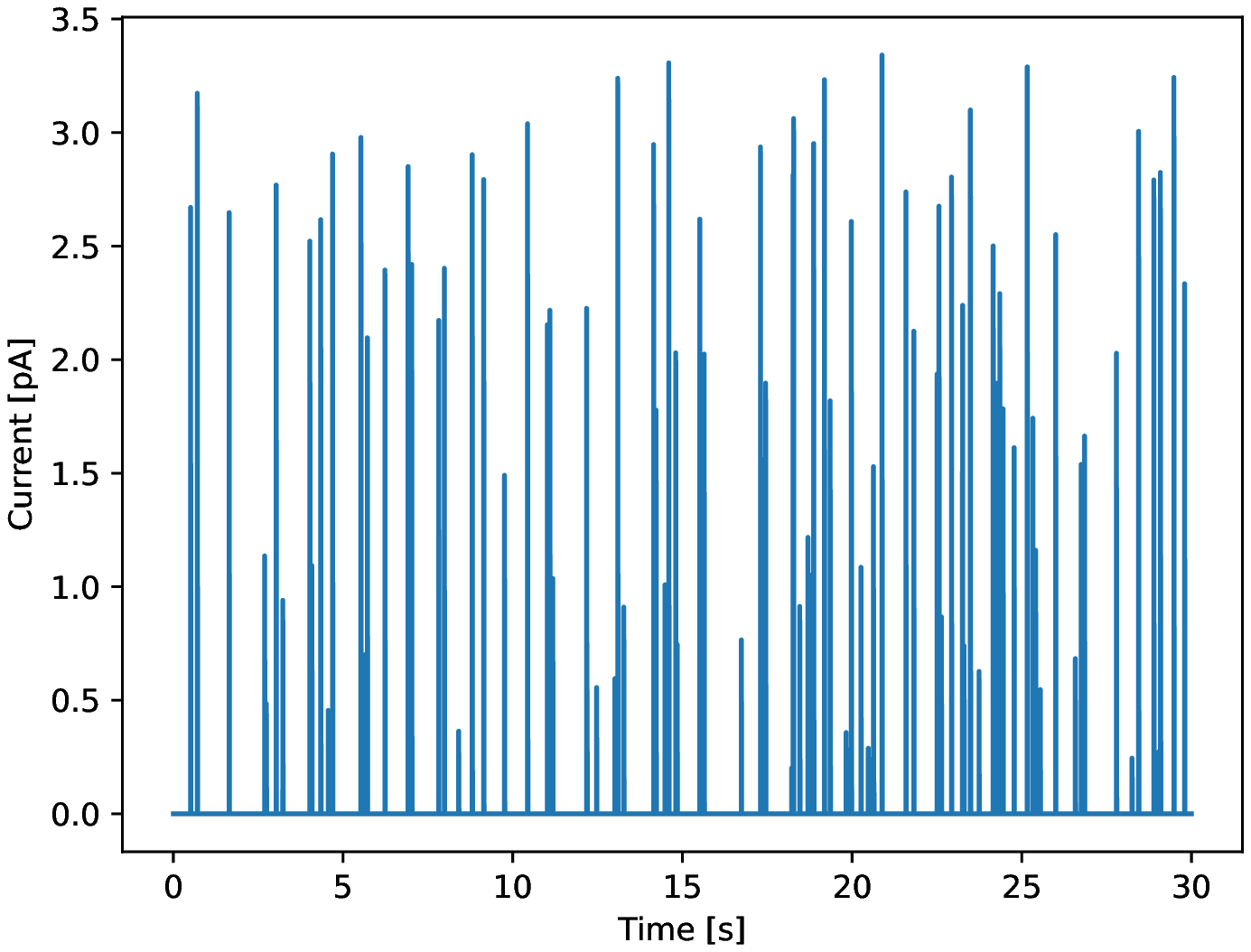}
        \caption[]%
        {{\small Spike trains with Cl$^-$ stimulation}}    
        \label{fig:cl}
    \end{subfigure}
    \begin{subfigure}[b]{0.3\textwidth}
        \centering
		\includegraphics[scale=0.3]{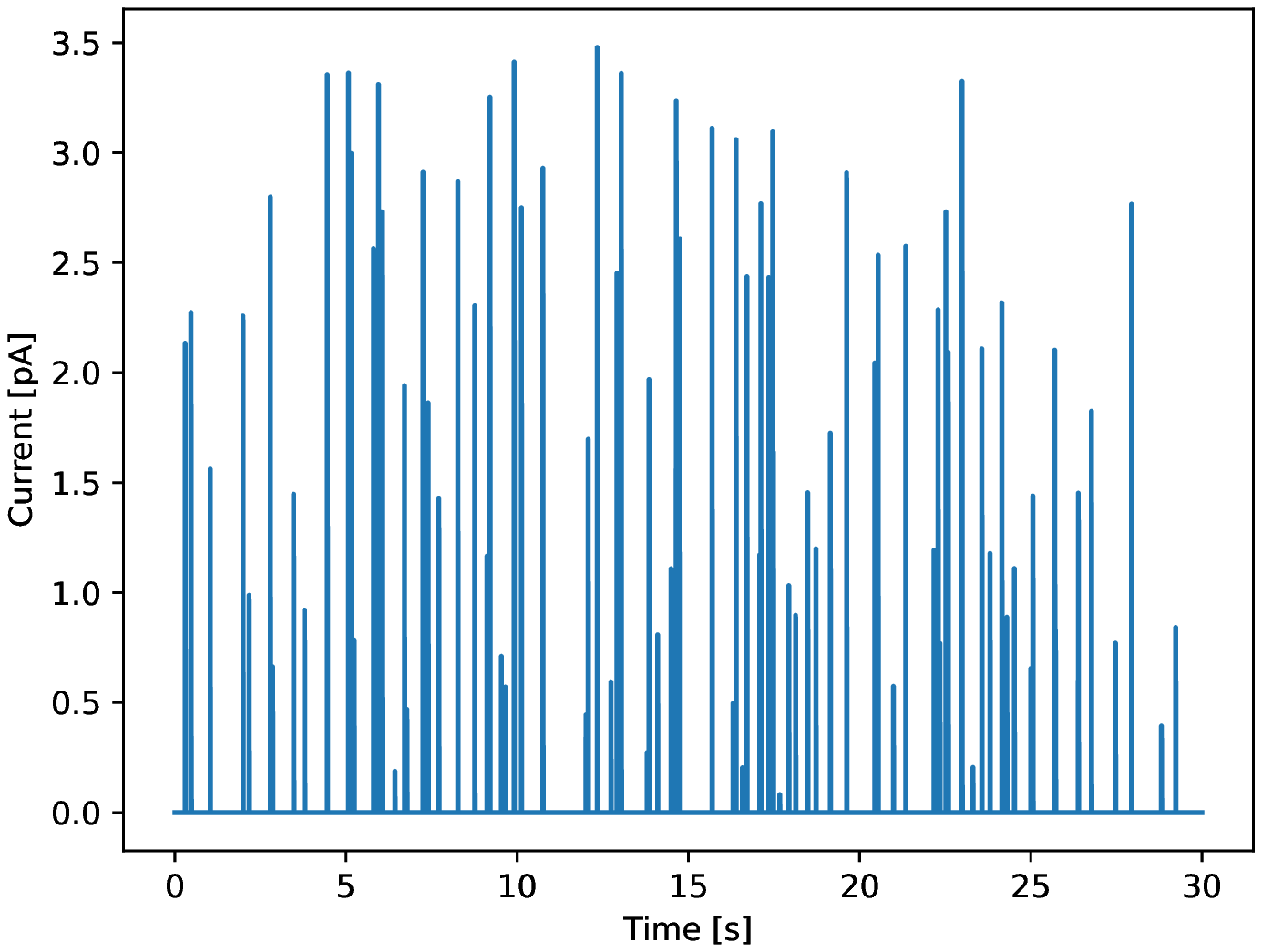}
        \caption[]%
        {{\small Spike trains with ClO$_4^-$ stimulation}}    
        \label{fig:clo4}
    \end{subfigure}
    \begin{subfigure}[b]{0.3\textwidth}
        \centering
		\includegraphics[scale=0.3]{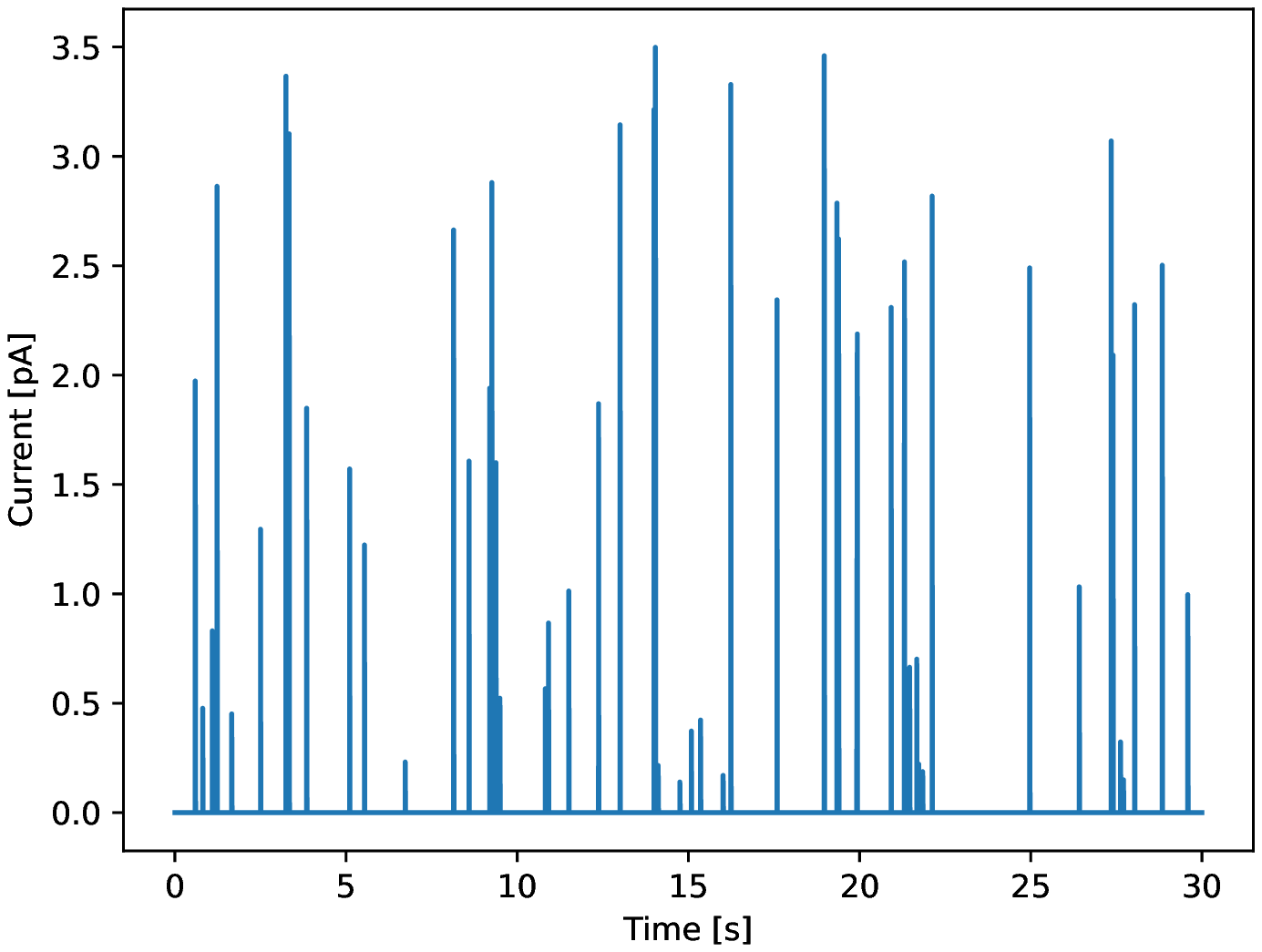}
        \caption[]%
        {{\small Spike trains with NO$_3^-$ stimulation}}    
        \label{fig:no3}
    \end{subfigure}
    \begin{subfigure}[b]{0.3\textwidth}
        \centering
		\includegraphics[scale=0.3]{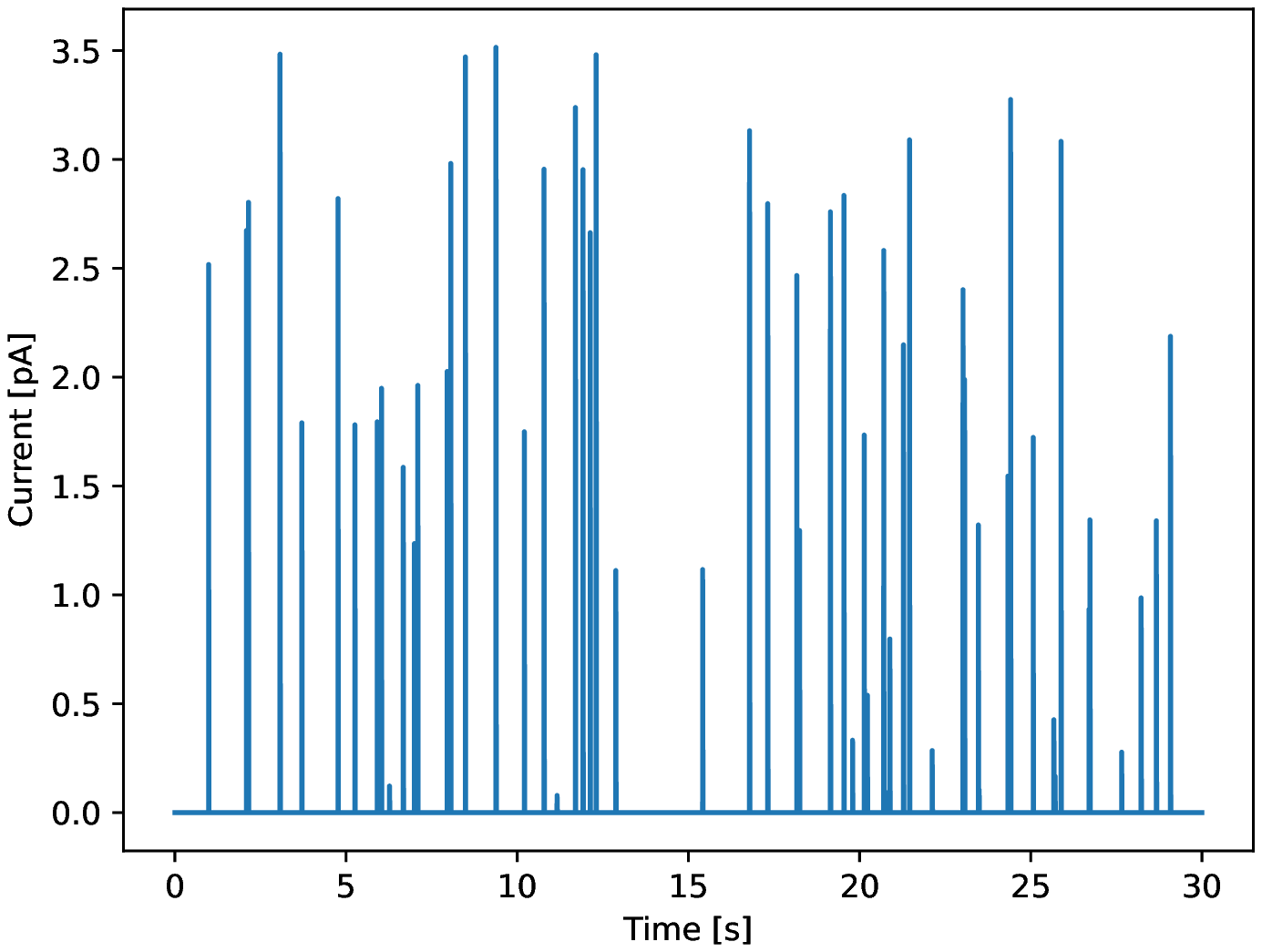}
        \caption[]%
        {{\small Spike trains with SCN$^-$ stimulation}}    
        \label{fig:scn}
    \end{subfigure}
    \caption{Amperometric traces of chromaffin cells under different ion stimulations obtained through SCA experiments}
    \label{fig:traces}
\end{figure}
\subsubsection*{DMSO Dataset}
Bovine chromaffin cells were isolated from the adrenal medulla by enzymatic digestion and the cells were kept at $37^\circ C$. Electrochemical recordings from single cells were performed on an inverted microscope, in a Faraday cage. The working electrode was held at $+ 700$ mV versus an Ag/AgCl reference electrode and the output was filtered at $2$ kHz by using a Bessel filter. For cytometry recording, the tip of the nanoelectrode was inserted through the cell membrane with a patch-clamp micromanipulator. For exocytosis experiments, the nanotip electrode was positioned on top of the cell. Each cell was stimulated once with $2$ mM $\mathrm{Ba}^{2+}$ for $5$ seconds through the micropipette coupled to a microinjection system. 

\begin{figure}[!htbp]
    \centering
\begin{subfigure}[b]{0.3\textwidth}
	\centering
	\includegraphics[scale=0.3]{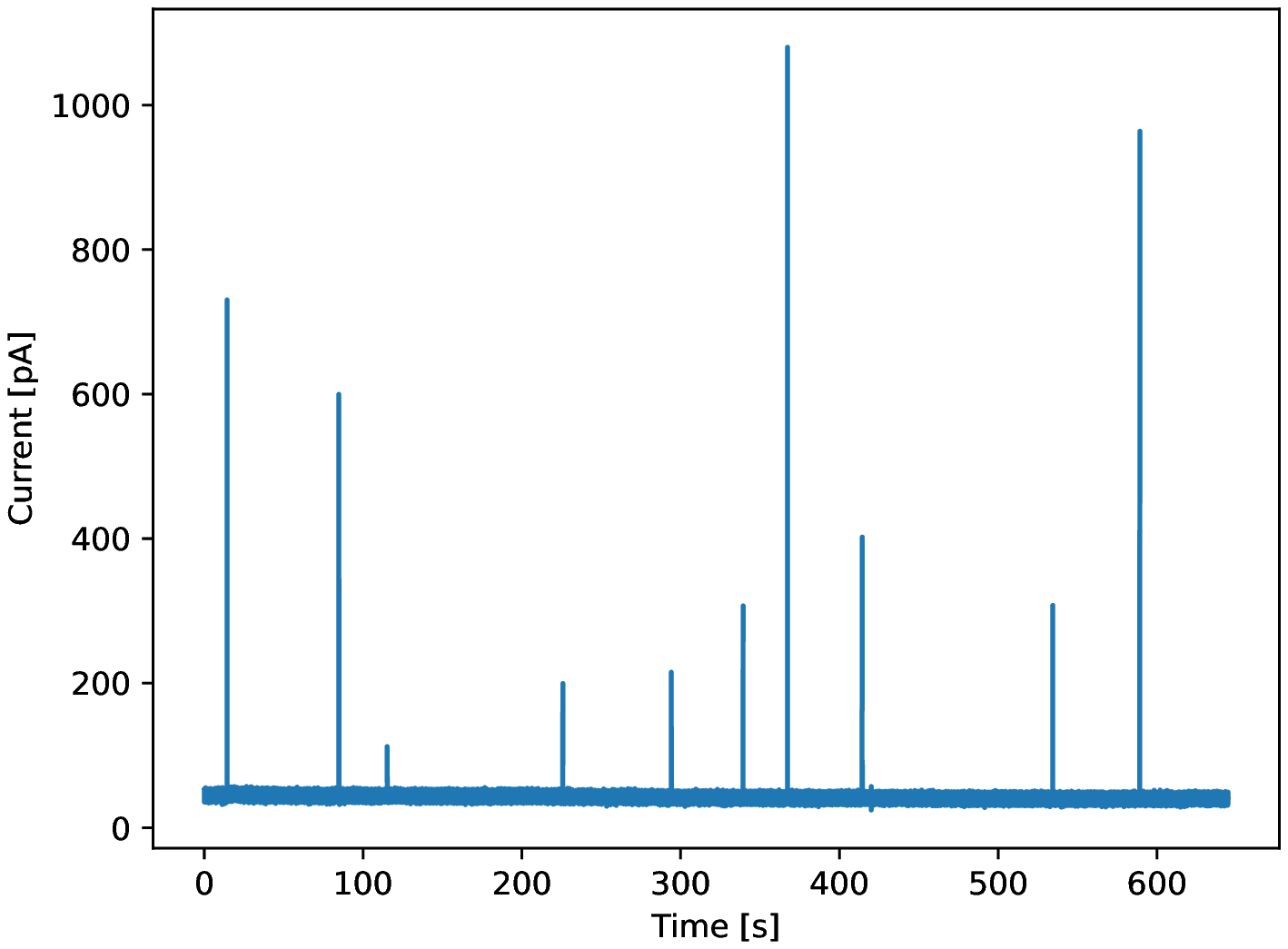}
	\caption[]%
	{{\small Spike trains without DMSO incubation}}    
	\label{fig:dmso_control}
\end{subfigure}
\begin{subfigure}[b]{0.3\textwidth}
	\centering
	\includegraphics[scale=0.3]{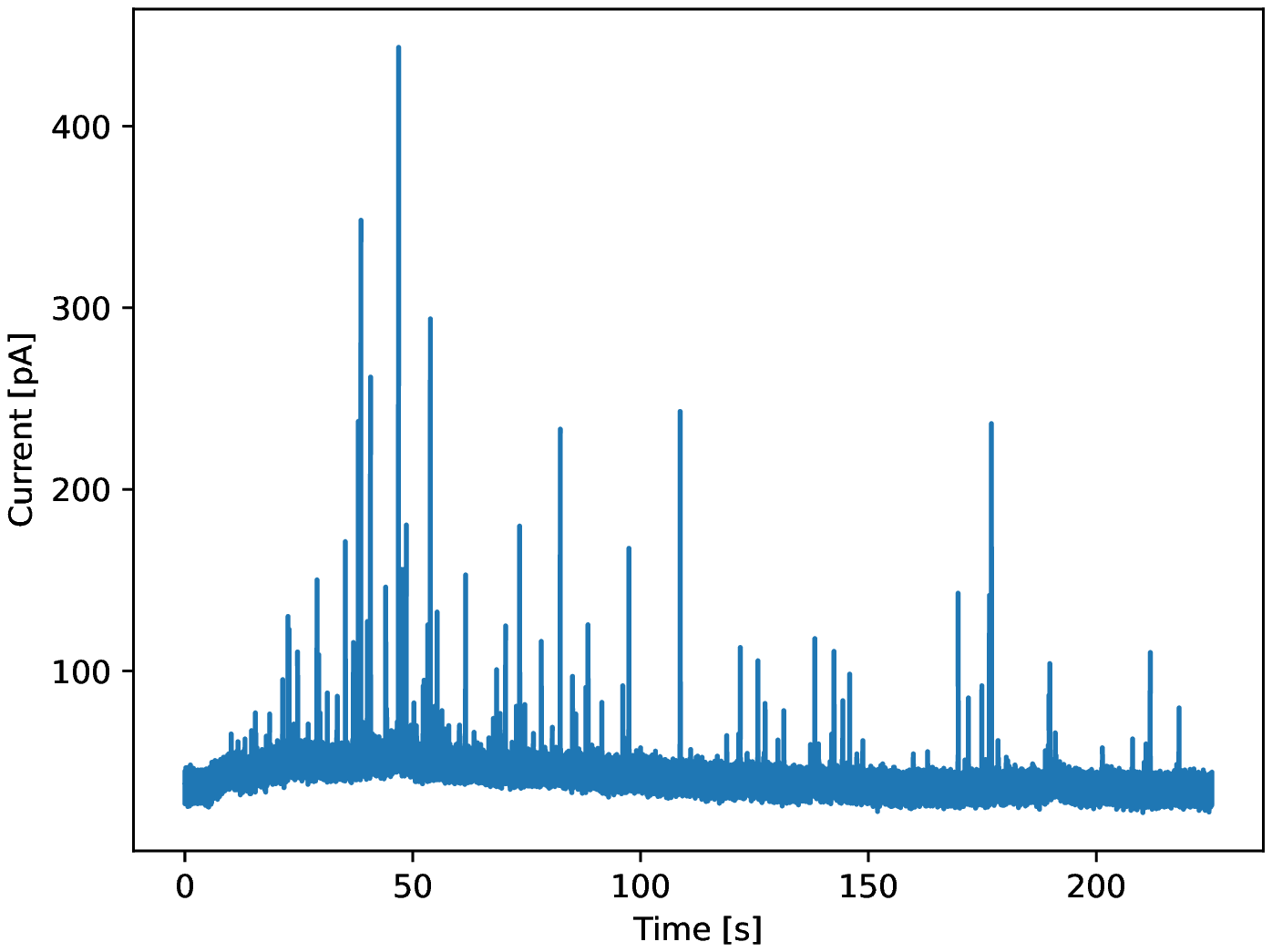}
	\caption[]%
	{{\small Spike trains with DMSO incubation}}    
	\label{fig:dmso}
\end{subfigure}
\caption{Amperometric traces of chromaffin cells under control conditions (no DMSO incubation) and with DMSO incubation obtained through IVIEC experiments}
\label{fig:traces_dmso}
\end{figure}
\subsubsection*{Electrodes Dataset}

\begin{figure}[!htbp]
\centering
\begin{subfigure}[b]{0.4\textwidth}
\includegraphics[scale=0.3]{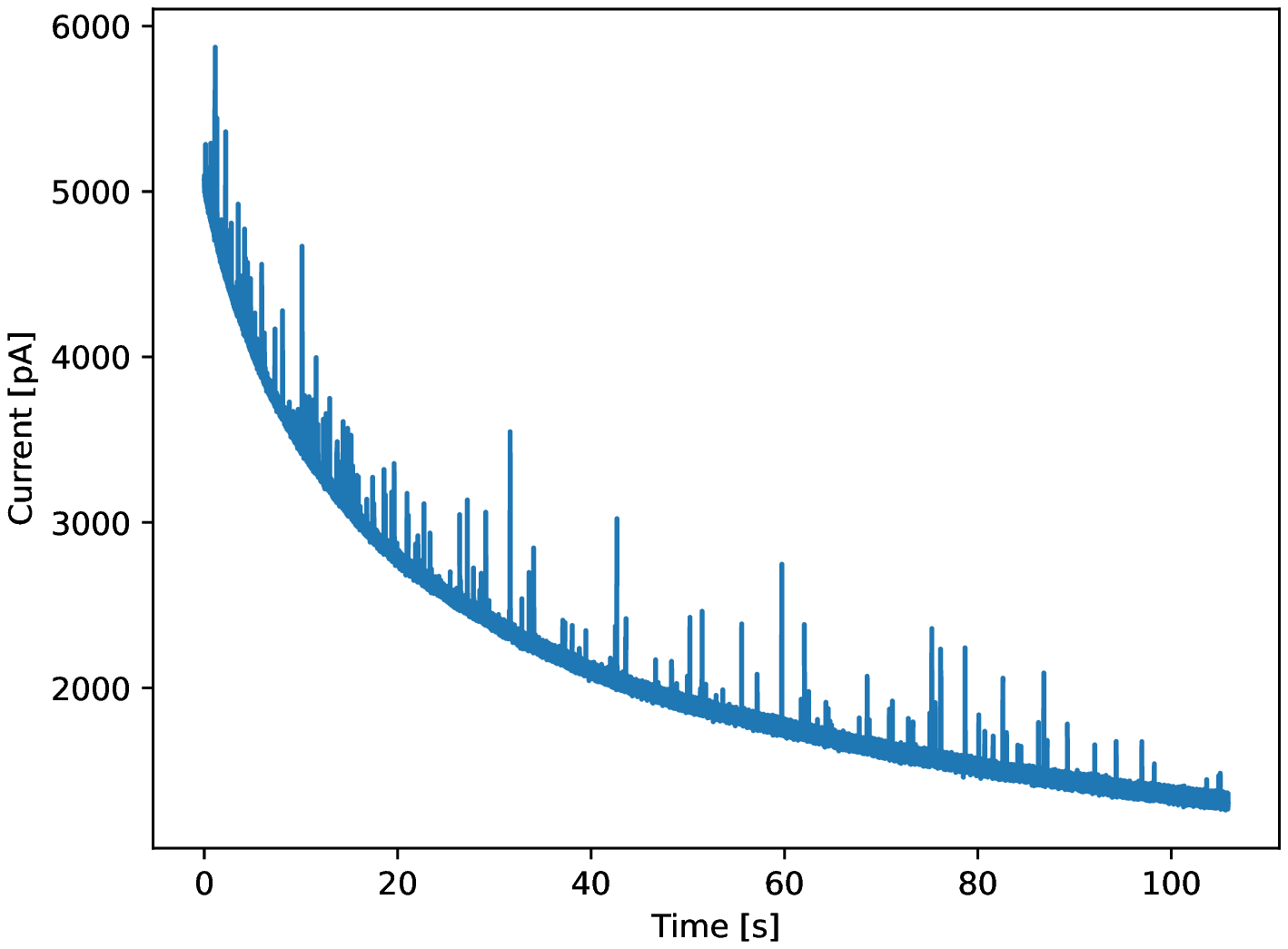}
\end{subfigure}
\begin{subfigure}[b]{0.4\textwidth}
\includegraphics[scale=0.3]{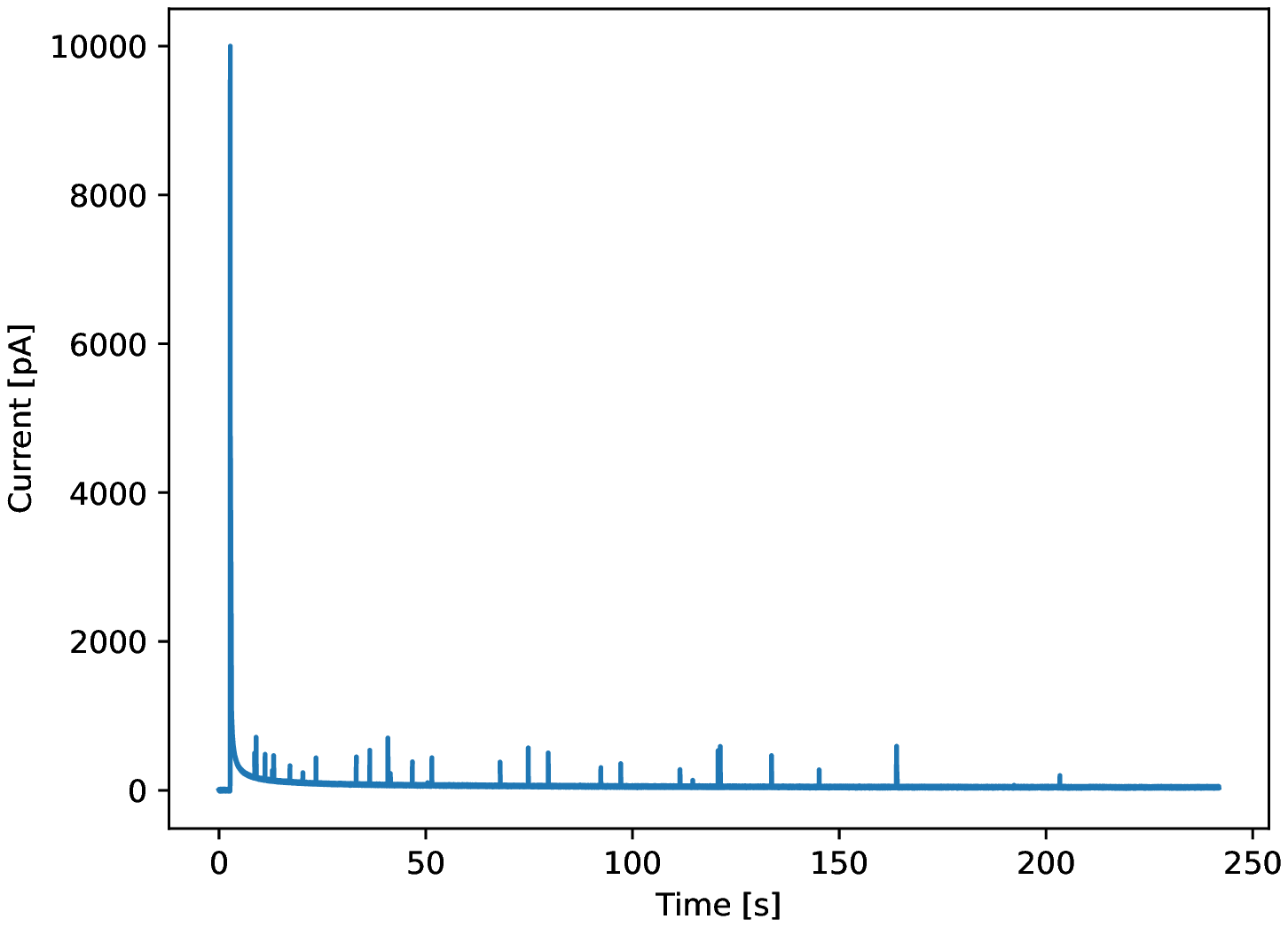}
\end{subfigure}\\
\begin{subfigure}[b]{0.4\textwidth}
\includegraphics[scale=0.3]{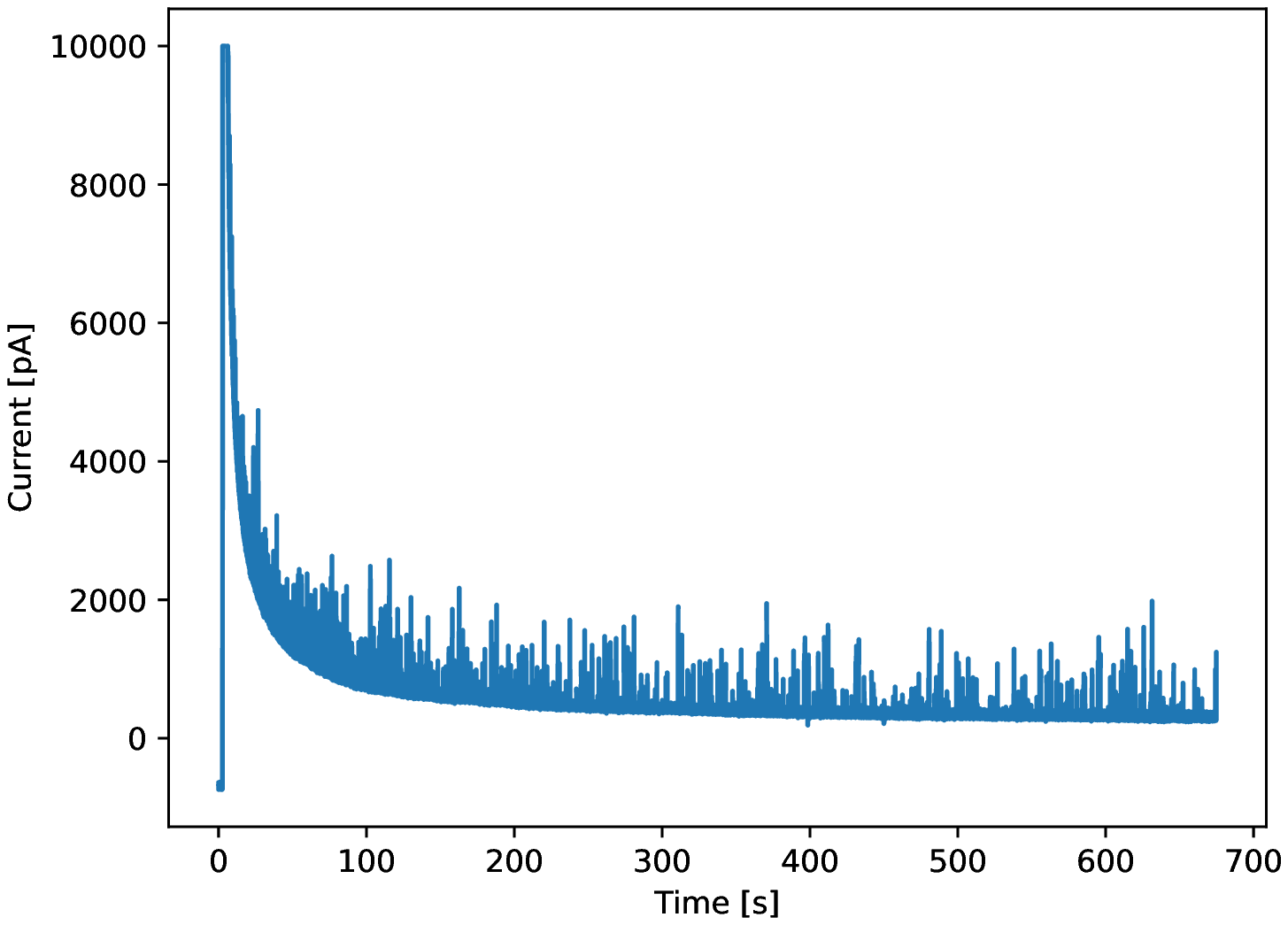}
\end{subfigure}
\caption{Amperometric traces of isolated chromaffin vesicles recorded at Au (top left), Pt (bottom) and Carbon (top right) disk microelectrodes at E = $+700$ mV vs Ag/AgCl obtained through VIEC experiments.}
\end{figure}

Bovine adrenal glands were obtained from a local slaughterhouse and transported in a cold Locke’s buffer. Glands were trimmed of surrounding fat and rinsed through adrenal vein with Locke's solution. The medulla was detached from the cortex with a scalpel and then mechanically homogenized in ice-cold homogenizing buffer. The homogenate was centrifuged at $1000\times$g for $10$ minutes to eliminate non-lysed cells and cell debris. After that, the supernatant was subsequently centrifuged at $10000\times$g for $20$ minutes to pellet vesicles. All centrifugation was performed at $4^{\circ}$C. The final pellet of chromaffin vesicles was resuspended and diluted in homogenizing buffer for VIEC measurements.

For the VIEC experiments, the electrodes were first dipped in a vesicle suspension for $30$ minutes at $4^{\circ}$C and then placed in homogenizing buffer for $20$ minutes at $37^{\circ}$C for experimental recording. During the measurements, a constant potential of $+700$ mV vs. Ag/AgCl reference electrode was applied to the working electrode using a low current potentiostat (Axopatch $200$B, Molecular Devices, Sunnyvale, CA, U.S.A). The signal output was filtered at $2$ kHz using a $4$-pole Bessel filter and digitized at $10$ kHz using a Digidata model $1440$A and Axoscope $10.3$ software (Axon Instruments Inc., Sunnyvale, CA, U.S.A.). 

For preparation of microelectrodes, a carbon fiber with $33$ $\mu m$ diameter was aspirated into a borosilicate capillary ($1.2$ mm $O.$D., $0.69$ mm I.D., Sutter Instrument Co., Novato, CA, U.S.A.). The capillaries were subsequently pulled using a micropipette puller (Narishige Inc., London, U.K) and the carbon fiber was cut at the glass junction. The gap between the carbon fiber and glass was sealed by dipping the pulled tip in epoxy. The glued electrodes were placed in an oven at $100^{\circ}$C overnight to complete the sealing step. The sealed electrodes were beveled at $45^{\circ}$ angle (EG-$400$, Narishige Inc., London, U.K.). A similar procedure was utilized for gold and platinum disk microelectrode fabrication. Here, either a $1$-cm length of $125$-$\mathrm{\mu-m}$-diameter Au wire or $100$-$\mathrm{\mu-m}$-diameter Pt wire (Goodfellow, Cambridge Ltd. U.K.) that was connected to a longer piece of a conductive wire (silver wire, $10$ cm) using silver paste was inserted into the pulled capillary and similarly sealed with epoxy and beveled at a $45^{\circ}$ angle. 

\clearpage
\subsection*{Artificial Data Generation}
\label{subsec:art_data}

An artificial dataset to test our frequency analysis hypothesis for the Hofmeister series dataset was created in the following manner. First all time series in our artificial dataset are assigned a fixed length of $300,000$, \ie\ $30$ seconds recording time assuming $10$ kHz sampling frequency. For each spike train, the number of spikes is randomly selected between [$50,100$], and each such spike is assigned a random width depending on the artificial ion type, \ie\ artificial \textbf{Cl$^-$} in [$10,20$], artificial \textbf{Br$^-$} in [$20,30$], artificial \textbf{NO$_3^-$} in [$30,40$], artificial \textbf{ClO$_4-$} in [$40,50$],  artificial \textbf{SCN$^-$} in [$50,60$], meant to mimic the observations in the real Hofmeister series dataset. Similar to real spike shapes, the artificial spikes consist of a steep linear rising segment and an exponentially decaying part. Finally, FFT was applied to each spike train and the statistics were pooled as shown in  Figure \ref{fig:art_fmean}. For each category, $25$ time series were generated. 

\end{suppinfo}

\clearpage
\bibliography{references}

\providecommand{\latin}[1]{#1}
\makeatletter
\providecommand{\doi}
  {\begingroup\let\do\@makeother\dospecials
  \catcode`\{=1 \catcode`\}=2 \doi@aux}
\providecommand{\doi@aux}[1]{\endgroup\texttt{#1}}
\makeatother
\providecommand*\mcitethebibliography{\thebibliography}
\csname @ifundefined\endcsname{endmcitethebibliography}
  {\let\endmcitethebibliography\endthebibliography}{}
\begin{mcitethebibliography}{36}
\providecommand*\natexlab[1]{#1}
\providecommand*\mciteSetBstSublistMode[1]{}
\providecommand*\mciteSetBstMaxWidthForm[2]{}
\providecommand*\mciteBstWouldAddEndPuncttrue
  {\def\EndOfBibitem{\unskip.}}
\providecommand*\mciteBstWouldAddEndPunctfalse
  {\let\EndOfBibitem\relax}
\providecommand*\mciteSetBstMidEndSepPunct[3]{}
\providecommand*\mciteSetBstSublistLabelBeginEnd[3]{}
\providecommand*\EndOfBibitem{}
\mciteSetBstSublistMode{f}
\mciteSetBstMaxWidthForm{subitem}{(\alph{mcitesubitemcount})}
\mciteSetBstSublistLabelBeginEnd
  {\mcitemaxwidthsubitemform\space}
  {\relax}
  {\relax}

\bibitem[Wrenn(2003)]{Wrenn2003}
Wrenn,~K. {Past and present.} \emph{Annals of internal medicine} \textbf{2003},
  \emph{138}, 847\relax
\mciteBstWouldAddEndPuncttrue
\mciteSetBstMidEndSepPunct{\mcitedefaultmidpunct}
{\mcitedefaultendpunct}{\mcitedefaultseppunct}\relax
\EndOfBibitem
\bibitem[Liu \latin{et~al.}(2019)Liu, Tong, and Fang]{Liu2019}
Liu,~X.; Tong,~Y.; Fang,~P.~P. {Recent development in amperometric measurements
  of vesicular exocytosis}. \emph{TrAC - Trends in Analytical Chemistry}
  \textbf{2019}, \emph{113}, 13--24\relax
\mciteBstWouldAddEndPuncttrue
\mciteSetBstMidEndSepPunct{\mcitedefaultmidpunct}
{\mcitedefaultendpunct}{\mcitedefaultseppunct}\relax
\EndOfBibitem
\bibitem[Colliver \latin{et~al.}(2000)Colliver, Hess, Pothos, Sulzer, and
  Ewing]{Colliver2000}
Colliver,~T.~L.; Hess,~E.~J.; Pothos,~E.~N.; Sulzer,~D.; Ewing,~A.~G.
  {Quantitative and statistical analysis of the shape of amperometric spikes
  recorded from two populations of cells}. \textbf{2000}, \relax
\mciteBstWouldAddEndPunctfalse
\mciteSetBstMidEndSepPunct{\mcitedefaultmidpunct}
{}{\mcitedefaultseppunct}\relax
\EndOfBibitem
\bibitem[Segura \latin{et~al.}(2000)Segura, Brioso, G{\'{o}}mez, Machado, and
  Borges]{Segura2000}
Segura,~F.; Brioso,~M.~A.; G{\'{o}}mez,~J.~F.; Machado,~J.~D.; Borges,~R.
  {Automatic analysis for amperometrical recordings of exocytosis}.
  \emph{Journal of Neuroscience Methods} \textbf{2000}, \emph{103},
  151--156\relax
\mciteBstWouldAddEndPuncttrue
\mciteSetBstMidEndSepPunct{\mcitedefaultmidpunct}
{\mcitedefaultendpunct}{\mcitedefaultseppunct}\relax
\EndOfBibitem
\bibitem[Mosharov and Sulzer(2005)Mosharov, and Sulzer]{mosharov2005analysis}
Mosharov,~E.~V.; Sulzer,~D. {Analysis of exocytotic events recorded by
  amperometry}. \emph{Nature methods} \textbf{2005}, \emph{2}, 651--658\relax
\mciteBstWouldAddEndPuncttrue
\mciteSetBstMidEndSepPunct{\mcitedefaultmidpunct}
{\mcitedefaultendpunct}{\mcitedefaultseppunct}\relax
\EndOfBibitem
\bibitem[Lema{\^{i}}tre \latin{et~al.}(2014)Lema{\^{i}}tre, {Guille Collignon},
  and Amatore]{Lemaitre2014}
Lema{\^{i}}tre,~F.; {Guille Collignon},~M.; Amatore,~C. {Recent advances in
  Electrochemical Detection of Exocytosis}. \emph{Electrochimica Acta}
  \textbf{2014}, \emph{140}, 457--466\relax
\mciteBstWouldAddEndPuncttrue
\mciteSetBstMidEndSepPunct{\mcitedefaultmidpunct}
{\mcitedefaultendpunct}{\mcitedefaultseppunct}\relax
\EndOfBibitem
\bibitem[Wightman \latin{et~al.}(1991)Wightman, Jankowski, Kennedy, Kawagoe,
  Schroeder, Leszczyszyn, Near, Diliberto, and Viveros]{Wightman1991}
Wightman,~R.~M.; Jankowski,~J.~A.; Kennedy,~R.~T.; Kawagoe,~K.~T.;
  Schroeder,~T.~J.; Leszczyszyn,~D.~J.; Near,~J.~A.; Diliberto,~E.~J.;
  Viveros,~O.~H. Temporally Resolved Catecholamine Spikes Correspond to Single
  Vesicle Release from Individual Chromaffin Cells. \emph{Proc. Natl. Acad.
  Sci. U. S. A.} \textbf{1991}, \emph{88}, 10754--10758\relax
\mciteBstWouldAddEndPuncttrue
\mciteSetBstMidEndSepPunct{\mcitedefaultmidpunct}
{\mcitedefaultendpunct}{\mcitedefaultseppunct}\relax
\EndOfBibitem
\bibitem[Phan \latin{et~al.}(2017)Phan, Li, and Ewing]{Phan2017}
Phan,~N. T.~N.; Li,~X.; Ewing,~A.~G. Measuring Synaptic Vesicles Using Cellular
  Electrochemistry and Nanoscale Molecular Imaging. \emph{Nat. Rev. Chem.}
  \textbf{2017}, \emph{1}, 0048\relax
\mciteBstWouldAddEndPuncttrue
\mciteSetBstMidEndSepPunct{\mcitedefaultmidpunct}
{\mcitedefaultendpunct}{\mcitedefaultseppunct}\relax
\EndOfBibitem
\bibitem[Li \latin{et~al.}(2016)Li, Dunevall, and Ewing]{Li2016}
Li,~X.; Dunevall,~J.; Ewing,~A.~G. Quantitative Chemical Measurements of
  Vesicular Transmitters with Electrochemical Cytometry. \emph{Acc. Chem. Res.}
  \textbf{2016}, \emph{49}, 2347–2354\relax
\mciteBstWouldAddEndPuncttrue
\mciteSetBstMidEndSepPunct{\mcitedefaultmidpunct}
{\mcitedefaultendpunct}{\mcitedefaultseppunct}\relax
\EndOfBibitem
\bibitem[Dunevall \latin{et~al.}(2015)Dunevall, Fathali, Najafinobar, Lovric,
  Wigstr{\"{o}}m, Cans, and Ewing]{Dunevall2015}
Dunevall,~J.; Fathali,~H.; Najafinobar,~N.; Lovric,~J.; Wigstr{\"{o}}m,~J.;
  Cans,~A.~S.; Ewing,~A.~G. {Characterizing the Catecholamine Content of Single
  Mammalian Vesicles by Collision-Adsorption Events at an Electrode}.
  \emph{Journal of the American Chemical Society} \textbf{2015}, \emph{137},
  4344--4346\relax
\mciteBstWouldAddEndPuncttrue
\mciteSetBstMidEndSepPunct{\mcitedefaultmidpunct}
{\mcitedefaultendpunct}{\mcitedefaultseppunct}\relax
\EndOfBibitem
\bibitem[Li \latin{et~al.}(2015)Li, Majdi, Dunevall, Fathali, and
  Ewing]{Li2015}
Li,~X.; Majdi,~S.; Dunevall,~J.; Fathali,~H.; Ewing,~A.~G. {Quantitative
  Measurement of Transmitters in Individual Vesicles in the Cytoplasm of Single
  Cells with Nanotip Electrodes}. \emph{Angewandte Chemie - International
  Edition} \textbf{2015}, \emph{54}, 11978--11982\relax
\mciteBstWouldAddEndPuncttrue
\mciteSetBstMidEndSepPunct{\mcitedefaultmidpunct}
{\mcitedefaultendpunct}{\mcitedefaultseppunct}\relax
\EndOfBibitem
\bibitem[Ren \latin{et~al.}(2016)Ren, Mellander, Keighron, Cans, Kurczy, Svir,
  Oleinick, Amatore, and Ewing]{Ren2016}
Ren,~L.; Mellander,~L.~J.; Keighron,~J.; Cans,~A.; Kurczy,~M.~E.; Svir,~I.;
  Oleinick,~A.; Amatore,~C.; Ewing,~A.~G. The Evidence for Open and Closed
  Exocytosis as the Primary Release Mechanism. \emph{Q. Rev. Biophys}
  \textbf{2016}, \emph{49}, 1--27\relax
\mciteBstWouldAddEndPuncttrue
\mciteSetBstMidEndSepPunct{\mcitedefaultmidpunct}
{\mcitedefaultendpunct}{\mcitedefaultseppunct}\relax
\EndOfBibitem
\bibitem[Larsson \latin{et~al.}(2020)Larsson, Majdi, Oleinick, Svir, Dunevall,
  Amatore, and Ewing]{Larsson2020}
Larsson,~A.; Majdi,~S.; Oleinick,~A.; Svir,~I.; Dunevall,~J.; Amatore,~C.;
  Ewing,~A.~G. {Intracellular Electrochemical Nanomeasurements Reveal that
  Exocytosis of Molecules at Living Neurons is Subquantal and Complex}.
  \emph{Angewandte Chemie - International Edition} \textbf{2020}, \emph{59},
  6711--6714\relax
\mciteBstWouldAddEndPuncttrue
\mciteSetBstMidEndSepPunct{\mcitedefaultmidpunct}
{\mcitedefaultendpunct}{\mcitedefaultseppunct}\relax
\EndOfBibitem
\bibitem[Wang and Ewing(2021)Wang, and Ewing]{Wang2021}
Wang,~Y.; Ewing,~A. Electrochemical Quantification of Neurotransmitters in
  Single Live Cell Vesicles Shows Exocytosis Is Predominantly Partial.
  \emph{ChemBioChem} \textbf{2021}, \emph{22}, 807--813\relax
\mciteBstWouldAddEndPuncttrue
\mciteSetBstMidEndSepPunct{\mcitedefaultmidpunct}
{\mcitedefaultendpunct}{\mcitedefaultseppunct}\relax
\EndOfBibitem
\bibitem[{Steven M. Singer, Marc Y. Fink}(2019)]{Steven2019}
{Steven M. Singer, Marc Y. Fink},~V. V.~A. {Vesicle impact electrochemical
  cytometry compared to amperometric exocytosis measurements}. \emph{Physiology
  {\&} behavior} \textbf{2019}, \emph{176}, 139--148\relax
\mciteBstWouldAddEndPuncttrue
\mciteSetBstMidEndSepPunct{\mcitedefaultmidpunct}
{\mcitedefaultendpunct}{\mcitedefaultseppunct}\relax
\EndOfBibitem
\bibitem[Wacker and Witte(2013)Wacker, and Witte]{Wacker2013}
Wacker,~M.; Witte,~H. {Time-frequency techniques in biomedical signal analysis:
  A tutorial review of similarities and differences}. \emph{Methods of
  Information in Medicine} \textbf{2013}, \emph{52}, 279--296\relax
\mciteBstWouldAddEndPuncttrue
\mciteSetBstMidEndSepPunct{\mcitedefaultmidpunct}
{\mcitedefaultendpunct}{\mcitedefaultseppunct}\relax
\EndOfBibitem
\bibitem[Evanko(2005)]{Evanko2005}
Evanko,~D. {Primer: Spying on exocytosis with amperometry}. \emph{Nature
  Methods} \textbf{2005}, \emph{2}, 650\relax
\mciteBstWouldAddEndPuncttrue
\mciteSetBstMidEndSepPunct{\mcitedefaultmidpunct}
{\mcitedefaultendpunct}{\mcitedefaultseppunct}\relax
\EndOfBibitem
\bibitem[Cooley and Tukey(1965)Cooley, and Tukey]{cooleyandtuckey1965}
Cooley,~J.~W.; Tukey,~J.~W. An Algorithm for the Machine Calculation of Complex
  Fourier Series. \emph{Mathematics of Computation} \textbf{1965}, \emph{19},
  297--301\relax
\mciteBstWouldAddEndPuncttrue
\mciteSetBstMidEndSepPunct{\mcitedefaultmidpunct}
{\mcitedefaultendpunct}{\mcitedefaultseppunct}\relax
\EndOfBibitem
\bibitem[RN(1999)]{bracewell1999}
RN,~B. The Fourier transform and its applications. \emph{Mathematics of
  Computation} \textbf{1999}, \relax
\mciteBstWouldAddEndPunctfalse
\mciteSetBstMidEndSepPunct{\mcitedefaultmidpunct}
{}{\mcitedefaultseppunct}\relax
\EndOfBibitem
\bibitem[Brigham and Morrow(1967)Brigham, and Morrow]{Brigham1967}
Brigham,~E.~O.; Morrow,~R.~E. {The fast Fourier transform}. \emph{IEEE
  Spectrum} \textbf{1967}, \emph{4}, 63--70\relax
\mciteBstWouldAddEndPuncttrue
\mciteSetBstMidEndSepPunct{\mcitedefaultmidpunct}
{\mcitedefaultendpunct}{\mcitedefaultseppunct}\relax
\EndOfBibitem
\bibitem[Cohen(1989)]{cohen1989}
Cohen,~L. Time-frequency distributions-a review. \emph{Proceedings of the IEEE}
  \textbf{1989}, \emph{77}, 941--981\relax
\mciteBstWouldAddEndPuncttrue
\mciteSetBstMidEndSepPunct{\mcitedefaultmidpunct}
{\mcitedefaultendpunct}{\mcitedefaultseppunct}\relax
\EndOfBibitem
\bibitem[Cooley \latin{et~al.}(1967)Cooley, Lewis, and Welch]{cooley1967}
Cooley,~J.; Lewis,~P.; Welch,~P. Historical notes on the fast Fourier
  transform. \emph{Proceedings of the IEEE} \textbf{1967}, \emph{55},
  1675--1677\relax
\mciteBstWouldAddEndPuncttrue
\mciteSetBstMidEndSepPunct{\mcitedefaultmidpunct}
{\mcitedefaultendpunct}{\mcitedefaultseppunct}\relax
\EndOfBibitem
\bibitem[Coo(1969)]{Cooley1969}
{The Fast Fourier Transform and its Applications}. \emph{IEEE Transactions on
  Education} \textbf{1969}, \emph{12}, 27--34\relax
\mciteBstWouldAddEndPuncttrue
\mciteSetBstMidEndSepPunct{\mcitedefaultmidpunct}
{\mcitedefaultendpunct}{\mcitedefaultseppunct}\relax
\EndOfBibitem
\bibitem[Letelier and Weber(2000)Letelier, and Weber]{Letelier2000}
Letelier,~J.~C.; Weber,~P.~P. {Spike sorting based on discrete wavelet
  transform coefficients}. \emph{Journal of Neuroscience Methods}
  \textbf{2000}, \emph{101}, 93--106\relax
\mciteBstWouldAddEndPuncttrue
\mciteSetBstMidEndSepPunct{\mcitedefaultmidpunct}
{\mcitedefaultendpunct}{\mcitedefaultseppunct}\relax
\EndOfBibitem
\bibitem[Stankovic(1994)]{Stankovic1994}
Stankovic,~L. A Method for Time-Frequency Analysis. \textbf{1994}, \emph{42},
  0--4\relax
\mciteBstWouldAddEndPuncttrue
\mciteSetBstMidEndSepPunct{\mcitedefaultmidpunct}
{\mcitedefaultendpunct}{\mcitedefaultseppunct}\relax
\EndOfBibitem
\bibitem[L.~E.~Alsop(1966)]{alsopandnawroozi1966}
L.~E.~Alsop,~A. A.~N. Faster Fourier analysis. \emph{Journal of Geophysical
  Research} \textbf{1966}, \emph{71}, 5482--5483\relax
\mciteBstWouldAddEndPuncttrue
\mciteSetBstMidEndSepPunct{\mcitedefaultmidpunct}
{\mcitedefaultendpunct}{\mcitedefaultseppunct}\relax
\EndOfBibitem
\bibitem[Heitler(2007)]{Heitler2007}
Heitler,~W.~J. {DataView: A tutorial tool for data analysis. Template-based
  spike sorting and frequency analysis}. \emph{Journal of Undergraduate
  Neuroscience Education} \textbf{2007}, \emph{6}, 1--7\relax
\mciteBstWouldAddEndPuncttrue
\mciteSetBstMidEndSepPunct{\mcitedefaultmidpunct}
{\mcitedefaultendpunct}{\mcitedefaultseppunct}\relax
\EndOfBibitem
\bibitem[Mikheev(2015)]{Mikheev2015}
Mikheev,~P.~A. {Application of the fast Fourier transform to calculating pruned
  convolution}. \emph{Doklady Mathematics} \textbf{2015}, \emph{92},
  630--633\relax
\mciteBstWouldAddEndPuncttrue
\mciteSetBstMidEndSepPunct{\mcitedefaultmidpunct}
{\mcitedefaultendpunct}{\mcitedefaultseppunct}\relax
\EndOfBibitem
\bibitem[A.(1962)]{papoulis1962}
A.,~P. \emph{The Fourier Integral and its Applications}; New York: McGraw-Hill,
  1962\relax
\mciteBstWouldAddEndPuncttrue
\mciteSetBstMidEndSepPunct{\mcitedefaultmidpunct}
{\mcitedefaultendpunct}{\mcitedefaultseppunct}\relax
\EndOfBibitem
\bibitem[Tib(2013)]{Tibau2013}
{Identification of neuronal network properties from the spectral analysis of
  calcium imaging signals in neuronal cultures}. \emph{Frontiers in Neural
  Circuits} \textbf{2013}, \emph{7}, 1--16\relax
\mciteBstWouldAddEndPuncttrue
\mciteSetBstMidEndSepPunct{\mcitedefaultmidpunct}
{\mcitedefaultendpunct}{\mcitedefaultseppunct}\relax
\EndOfBibitem
\bibitem[Ruffinatti \latin{et~al.}(2011)Ruffinatti, Lovisolo, Distasi, Ariano,
  Erriquez, and Ferraro]{Ruffinatti2011}
Ruffinatti,~F.~A.; Lovisolo,~D.; Distasi,~C.; Ariano,~P.; Erriquez,~J.;
  Ferraro,~M. {Calcium signals : Analysis in time and frequency domains}.
  \emph{Journal of Neuroscience Methods} \textbf{2011}, \emph{199},
  310--320\relax
\mciteBstWouldAddEndPuncttrue
\mciteSetBstMidEndSepPunct{\mcitedefaultmidpunct}
{\mcitedefaultendpunct}{\mcitedefaultseppunct}\relax
\EndOfBibitem
\bibitem[He and Ewing()He, and Ewing]{He}
He,~X.; Ewing,~A.~G. {Counter Anions Alter Exocytotic Parameters along the
  Counter Anions Alter Exocytotic Parameters along the Hofmeister Series}.
  \relax
\mciteBstWouldAddEndPunctfalse
\mciteSetBstMidEndSepPunct{\mcitedefaultmidpunct}
{}{\mcitedefaultseppunct}\relax
\EndOfBibitem
\bibitem[Majdi \latin{et~al.}(2017)Majdi, Najafinobar, Dunevall, Lovric, and
  Ewing]{Majdi2017}
Majdi,~S.; Najafinobar,~N.; Dunevall,~J.; Lovric,~J.; Ewing,~A.~G. {DMSO
  Chemically Alters Cell Membranes to Slow Exocytosis and Increase the Fraction
  of Partial Transmitter Released}. \emph{ChemBioChem} \textbf{2017},
  \emph{18}, 1898--1902\relax
\mciteBstWouldAddEndPuncttrue
\mciteSetBstMidEndSepPunct{\mcitedefaultmidpunct}
{\mcitedefaultendpunct}{\mcitedefaultseppunct}\relax
\EndOfBibitem
\bibitem[Coifman \latin{et~al.}(1992)Coifman, Meyer, and
  Wickerhauser]{Coifman92}
Coifman,~R.~R.; Meyer,~Y.; Wickerhauser,~V. Wavelet Analysis and Signal
  Processing. In Wavelets and their Applications. 1992; pp 153--178\relax
\mciteBstWouldAddEndPuncttrue
\mciteSetBstMidEndSepPunct{\mcitedefaultmidpunct}
{\mcitedefaultendpunct}{\mcitedefaultseppunct}\relax
\EndOfBibitem
\bibitem[Pavlov \latin{et~al.}(2012)Pavlov, Hramov, Koronovskii, Sitnikova,
  Makarov, and Ovchinnikov]{Pavlov_2012}
Pavlov,~A.~N.; Hramov,~A.~E.; Koronovskii,~A.~A.; Sitnikova,~E.~Y.;
  Makarov,~V.~A.; Ovchinnikov,~A.~A. Wavelet analysis in neurodynamics.
  \emph{Physics-Uspekhi} \textbf{2012}, \emph{55}, 845--875\relax
\mciteBstWouldAddEndPuncttrue
\mciteSetBstMidEndSepPunct{\mcitedefaultmidpunct}
{\mcitedefaultendpunct}{\mcitedefaultseppunct}\relax
\EndOfBibitem
\end{mcitethebibliography}
\end{document}